\documentclass[fleqn,usenatbib]{mnras}

\usepackage{newtxtext,newtxmath}

\usepackage[T1]{fontenc}
\usepackage{ae,aecompl}

\usepackage{graphicx}	
\usepackage{amsmath}	
\usepackage{amssymb}	
\usepackage{subcaption}
\usepackage{pifont}
\usepackage[table]{xcolor} 

\providecommand{\e}[1]{\ensuremath{\times 10^{#1}}}
\newcommand{\cmark}{\ding{51}}%
\newcommand{\xmark}{\ding{55}}%

\font\btt=rm-lmtk10 at 18truept

\title[Calibrating star formation and feedback]{Calibration of a star formation and feedback model for cosmological simulations with {\btt Enzo}}

\author[B.K. Oh et al.]{
Boon Kiat Oh,$^{1}$\thanks{E-mail: bkoh@roe.ac.uk}
Britton D. Smith,$^{1, 2}$
John A. Peacock,$^{1}$
and Sadegh Khochfar$^{1}$
\\
$^{1}$Institute for Astronomy, University of Edinburgh, Royal Observatory, Edinburgh EH9 3HJ, United Kingdom\\
$^{2}$San Diego Supercomputer Center, University of California, San Diego, 10100 Hopkins Drive, La Jolla, CA 92093\\
}

\pubyear{2019}

\begin{document}
\label{firstpage}
\pagerange{\pageref{firstpage}--\pageref{lastpage}}
\maketitle

\begin{abstract}
	We present results from seventy-one zoom simulations of a Milky Way-sized (MW) halo, exploring the parameter space for a widely-used star formation and feedback model in the {\tt Enzo} simulation code. We propose a novel way to match observations, using functional fits to the observed baryon makeup over a wide range of halo masses. The model MW galaxy is calibrated using three parameters: the star formation efficiency $\left(f_*\right)$, the efficiency of thermal energy from stellar feedback $\left(\epsilon\right)$ and the region into which feedback is injected $\left(r\ {\rm and}\ s\right)$. We find that changing the amount of feedback energy affects the baryon content most significantly. We then identify two sets of feedback parameter values that are both able to reproduce the baryonic properties for haloes between $10^{10}\,\mathrm{M_\odot}$ and $10^{12}\,\mathrm{M_\odot}$. We can potentially improve the agreement by incorporating more parameters or physics. If we choose to focus on one property at a time, we can obtain a more realistic halo baryon makeup. We show that the employed feedback prescription is insensitive to dark matter mass resolution between $10^5\,{\rm M_\odot}$ and $10^7\,{\rm M_\odot}$. Contrasting both star formation criteria and the corresponding combination of optimal feedback parameters, we also highlight that feedback is self-consistent: to match the same baryonic properties, with a relatively higher gas to stars conversion efficiency, the feedback strength required is lower, and vice versa. Lastly, we demonstrate that chaotic variance in the code can cause deviations of approximately 10\% and 25\% in the stellar and baryon mass in simulations evolved from identical initial conditions. 
\end{abstract}




\section{Introduction}

The large-scale structure of the universe can be understood quite precisely by considering models that consist purely of dark matter. Numerical simulations of structure formation in such models have been performed with high accuracy and progressively higher resolution and larger box size \citep{1985ApJS...57..241E, 1999ApJ...524L..19M,2008MNRAS.391.1685S,2008Natur.454..735D,2011ApJ...740..102K}. But on the baryonic side, limitations in numerical resolution mean that several baryonic processes are not simulated from first principles. These processes include fundamental phenomena of the transformation of cold gas to stars, feedback from the energy released by stars, supernovae and massive black holes. Such effects are implemented using a subgrid approach in cosmological hydrodynamical simulations \citep{2003MNRAS.339..289S, 2010Natur.463..203G, 2013ApJ...770...25A, 2019MNRAS.484.2632S}. If these analytical implementations are too simplistic, they risk being sensitive to poorly determined parameters, thus limiting their capability to make robust predictions. Improving the accuracy of subgrid physics requires both a better understanding of physical processes and identification of their limitations.

{\parskip 2em} Feedback processes are essential in order to solve fundamental issues in numerical simulations such as the `overcooling problem' \citep{1991ApJ...367...45C, 1991ApJ...379...52W, 1992A&A...264..365B} and the `angular momentum problem' \citep{1991ApJ...377..365K, 1994MNRAS.267..401N, 2012ApJ...749..140H}. Overcooling results in the formation of too massive galaxies particularly in high-resolution simulations \citep{2001ApJ...552..473D}. Feedback is also important for shaping the density profile of dark matter haloes \citep{2012MNRAS.421.3464P, 2013MNRAS.432.1947M, 2014MNRAS.443..985D}. In addition to these issues of small-scale subgrid physics, cosmological simulations contain additional uncertainties. In the absence of feedback, \citet{2018arXiv180707084G} highlighted differences in the properties of galaxies induced by very slight changes in the initial positions of dark matter particles. Even if a galaxy is evolved from identical initial conditions, the simulation code can introduce variances which result in fluctuations in the simulated properties between repetitions of the same simulation \citep{2019MNRAS.482.2244K}. The problem is alleviated by the self-regulating nature of feedback \citep{2019MNRAS.482.2244K} and highlights the need to understand the impact that subgrid implementations have on the resulting properties of galaxies in simulations.

{\parskip 2em} Feedback processes that inject energy into the gas are therefore integral to numerical simulations. For smaller mass haloes, the energy comes mainly from supernovae explosions. In contrast, for more massive ones, the main energy sources are active galactic nuclei (AGN) \citep{2007MNRAS.380..877S, 2009MNRAS.398...53B, 2011MNRAS.414..195T} and gravitational heating as a result of infalling clumps of matter \citep{2006MNRAS.368....2D, 2008ApJ...680...54K}. However, it is unclear how the energy should be distributed between generating motion and heating the gas. For supernova feedback alone, various techniques have been employed across different simulation codes \citep{2006MNRAS.373.1074S, 2006ApJ...650..560C, 2008A&A...477...79D, 2012MNRAS.426..140D, 2018MNRAS.478..302S}. Given the huge diversity in the method of implementation, it is not unusual to expect significantly different outcomes \citep{2000ApJ...545..728T, 2003MNRAS.339..289S, 2005MNRAS.363.1299O, 2006MNRAS.373.1265O, 2010MNRAS.402.1536S}, and variation in feedback effects is the most significant source of uncertainty in a cosmological simulation. In particular, the role of resolution should be emphasised: the resolution in cosmological simulations is limited but feedback occurs on all scales, so rigorous numerical convergence cannot be expected. The subgrid parameterisation, or at least the subgrid parameter values, will need to change according to the resolution in order to match calibrating observations, and there is no guarantee that all predicted galaxy properties will then be independent of resolution.

{\parskip 2em} To reproduce a realistic picture of the observed universe, there is thus a need to calibrate the parameters of the appropriate subgrid routines \citep{2015MNRAS.446..521S}. These are adjusted to match specific observational properties of the galaxy population. By matching related properties, the simulation can then be used to answer a wide range of questions. For example, the feedback implementation in the `Evolution and Assembly of GaLaxies and their Environments' (EAGLE) simulation project is calibrated to reproduce the observed $z=0.1$ galaxy stellar mass function (GSMF), the relation between the mass of galaxies and their central black holes and realistic galaxy sizes \citep{2015MNRAS.446..521S}. The Illustris group calibrate their parameters to match various observational scaling relations and galaxy properties at low and intermediate redshifts \citep{2014MNRAS.444.1518V}. Despite the calibrations, there are shortcomings in each simulation. For example, Illustris recognised that the decrease of their simulated cosmic star formation rate density was too slow, leading to an update in their feedback prescription, resulting in the introduction of IllustrisTNG \citep{2018MNRAS.473.4077P}.

{\parskip 2em} In contrast to these
full cosmologically representative box simulations, zoom simulations focus computational resources on smaller volumes \citep{2008MNRAS.391.1685S, 2016ApJ...818...10G, 2015MNRAS.454...83W}. In particular, \citet{2015MNRAS.454...83W} studied a halo mass range from dwarf masses ($5\e{9}\,\mathrm{M_\odot}$) to Milky Way (MW) masses ($2\e{12}\,\mathrm{M_\odot}$). They included baryonic processes and were able to reproduce the stellar to halo mass relation from abundance matching \citep{2013ApJ...770...57B, 2013MNRAS.428.3121M, 2013ApJ...764L..31K} across a wide range of redshifts. However, they did not account for the mass of gas remaining in the haloes, and this is an important issue for the present analysis.

{\parskip 2em} In this paper, we use zoom simulations of MW haloes in an attempt to quantify the stellar and gas mass present in such a halo at $z=0$. In particular, we examine the degree of calibration allowed by the model introduced by \citet{1992ApJ...399L.113C}. Although newer models are available, the Cen \& Ostriker model remains as one of the most highly-used models in {\tt Enzo} simulations. We calibrate our parameters governing star formation and feedback via a comparison with the inventory of baryonic and gravitating masses of cosmic structures presented in \citet{2010ApJ...708L..14M}, in particular the mass fraction of baryons in the halo and the conversion efficiency of gas into stars.  Not only is this the first suite of simulations using these observables for feedback calibration, it tests how well the \citet{1992ApJ...399L.113C} model can be calibrated. 

{\parskip 2em} This paper is structured as follows. Section \ref{simsetup} describes the generation of initial conditions used in the simulations, the code, and setup used to evolve them. Also, we describe the parameters used for calibration and analysis tools used to extract and analyse the results. Section \ref{observations} presents the properties from \citet{2010ApJ...708L..14M} that we attempt to match, along with the observational fit of the Kennicutt--Schmidt relation \citep{2007ApJ...671..333K}. Section \ref{results} describes the results from various simulations: effects of single parameter variation, calibration of parameters to results from \citet{2010ApJ...708L..14M} and performance of the simulations to match other constraints. Lastly, the results are summarised and discussed in Section \ref{discussion}.

\section{Simulation Setup and Analysis}\label{simsetup}

This section provides an overview of the simulation setup and the associated subgrid physics. In particular, the focus is on a MW-sized halo, at which mass scale we expect that AGN feedback will be subdominant \citep{2006MNRAS.370..645B, 2010ApJ...717..379B, 2014IAUS..303..354S}. The main parameters investigated will thus be related to star formation efficiency and supernova feedback, and one aim of this investigation is indeed to see to what extent we can reproduce the baryonic properties of the MW using solely these ingredients. As described by \citet{2015MNRAS.450.1937C}, the resulting baryonic properties of the halo are very sensitive to the variations of feedback parameter values. Therefore, a detailed explanation of the role of each parameter in the physical model is necessary.

{\parskip 2em} The cosmological parameters in this suite of simulations are taken from WMAP-9 \citep{2013ApJS..208...20B}. The key parameters are $\Omega_m=0.285$, $\Omega_\Lambda=0.715$, $\Omega_b=0.0461$, $h = 0.695$ and $\sigma_8=0.828$ with the usual definitions. With these parameters, we generate initial conditions with MUlti-scale Initial Conditions ({\tt MUSIC}) for cosmological simulations \citep{2011MNRAS.415.2101H}. We derive all zoom simulations from the parent simulation with a volume of $L=100\,h^{-1}\,\mathrm{cMpc}$ with $256^3$ particles. 

{\parskip 2em} The simulation is evolved using {\tt Enzo}, an adaptive mesh-refinement (AMR) code \citep{2014ApJS..211...19B}. {\tt Enzo} uses a block-structured AMR framework \citep{1989JCoPh..82...64B} to solve the equations of hydrodynamics in an Eulerian frame using multiple solvers. In the simulations presented here, we use the {\tt ZEUS} \citep{1992ApJS...80..753S} hydro solver in combination with an N-body adaptive particle-mesh gravity solver \citep{1985ApJS...57..241E}. Parameter space exploration is performed mainly on the star formation and feedback routines; the results of this exploration will be outlined extensively in Section \ref{sfmod} and \ref{fbpara}. Lastly, the chemistry and cooling processes are handled by the {\tt Grackle} library \citep{2017MNRAS.466.2217S}. We use the equilibrium cooling mode from {\tt Grackle}, which utilises the tabulated cooling rates derived from the photoionisation code CLOUDY \citep{2013RMxAA..49..137F} together with the UV background radiation given by \citet{2012ApJ...746..125H}. 

{\parskip 2em} The MW-sized halo is initially identified from a dark matter only parent simulation through its merger history and final dark matter halo mass. It is isolated, has not experienced a major merger in its merger history since at least $z=2$ and has a final mass of approximately $10^{12}\,\mathrm{M_\odot}$. The particles within a high-resolution region, typically larger than the virial radius, then undergo additional levels of refinement in mass while the region's spatial resolution is increased. Each nested level is equivalent to an increase in spatial and mass resolution by a factor of two and eight, respectively. Contamination occurs if larger mass particles cross the region of interest \citep{2014MNRAS.437.1894O}. In our simulations, we define a high-resolution region of three virial radii from the centre of the halo to carry out the refinement \citep{2018MNRAS.478..548S} as a preventive measure. We use three nested levels, giving an effective resolution of $2048^3$ particles or a nested dark matter particle mass of $1.104 \times 10^7\,\mathrm{M_\odot}$. This nested simulation is evolved with an additional five levels of AMR which is only allowed around particles within the high-resolution region, resulting in a maximum resolution of eight levels of spatial refinement or $2.196$ comoving kpc $({\rm ckpc})$. This simulation setup is similar to that presented by \citet{2019ApJ...873..129P} and \citet{2018arXiv181112410H}.

{\parskip 2em} From the high-resolution region of the MW halo, we identify an additional smaller halo with a mass of approximately $10^{10}\,\mathrm{M_\odot}$. We then run a separate simulation zooming in only on this halo with two additional levels of initial nesting. The purpose of this smaller halo is to test the universality of the optimal feedback parameters from the MW zoom simulation. Due to the additional nesting levels, the dwarf is made up of approximately the same number of dark matter particles as in the MW halo. The increased mass resolution translates into an effective resolution of $8192^3$ particles or a nested dark matter particle mass of $1.715\e{5}\,\mathrm{M_\odot}$. Because of the additional nested levels, we reduce the number of AMR levels to three, maintaining a constant maximum spatial resolution of $2.196\,{\rm ckpc}$.

\subsection{Star formation parameters}\label{sfmod}

This paper employs the model described by \citet{1992ApJ...399L.113C} with modifications for the purpose of calibration. This model is one of the most commonly used in {\tt Enzo}. The conditions required for star formation in a cell include:

\begin{enumerate}
	\item No further refinement within the cell
	\item Gas density greater than a threshold density: $\rho_\mathrm{gas} > \rho_\mathrm{threshold}$
	\item Convergent flow: $\boldsymbol{\nabla} \cdot {\bf v} < 0$
	\item Cooling time less than a dynamical time: $t_\mathrm{cool} < t_\mathrm{dyn}$
	\item Gas mass larger than the Jeans mass: $m_\mathrm{gas} > m_\mathrm{jeans}$
	\item Star particle mass is greater than a threshold mass
\end{enumerate}

{\parskip 2em} If all the conditions are fulfilled, the algorithm generates a `star particle' within the grid cell with a mass 
\begin{equation}\label{eq:mstar}
m_* = m_{\mathrm{gas}} \times \frac{\Delta t}{t_{\mathrm{dyn}}} \times f_*,
\end{equation}where $m_{\mathrm{gas}}$ is the gas mass in the cell, $\Delta t$ is the timestep, $t_{\mathrm{dyn}}$ is the dynamical time and $f_*$ is a dimensionless efficiency factor. The mass of the generated star particle is compared to a user-defined minimum star particle mass. If the mass exceeds the threshold, a star particle will be created. It is positioned in the centre of the cell and possesses the same peculiar velocity as the gas in the cell. It is treated dynamically as all other particles. An equivalent mass of gas to that of the star particle is then removed from the cell to ensure mass conservation. 

To calibrate the simulation, certain aspects of the star formation criteria are modified. These include the Jeans instability check, time dependence of star formation, threshold stellar mass and the value of $f_*$. The following sections will explain the role that each parameter plays: they are organized in the order that each factor is used in the star formation condition check.  

\subsubsection{Jeans instability check}\label{usejeansmass}
In item (v) of the list of conditions in Section \ref{sfmod}, the creation of star particles is only allowed when the gas mass exceeds the Jeans mass of the cell. This criterion is aimed at low resolution simulations that cannot resolve local Jeans masses. However, modern implementations with better resolution resolve such clouds with multiple cells at the star formation threshold density. When the spatial resolution of the simulation is high enough to resolve the Jeans length, this particular check in the star formation routine instead restricts star formation that can occur because an individual cell needs to wait until enough mass has accumulated within it.

\subsubsection{Minimum star particle mass}\label{minmass}

Once a cell fulfils all five conditions for star formation, the final barrier to star formation is the minimum mass of a star particle that will be inserted into the simulation. This threshold is explicitly designed to prevent the production of too many star particles, which can increase computational costs significantly. However, the inability to exceed this minimum star particle mass can lead to a build-up of potential star-forming gas in surrounding cells. This accumulation then reaches a point where a burst in star formation occurs.

\subsubsection{Timestep dependence of star formation}\label{timesf}

Two factors affect the mass of the star particle to be compared to the threshold value as seen in Equation \ref{eq:mstar}: $\Delta t / t_{\rm dyn}$ and $f_*$. They correspond to the timestep dependence of star formation and a conversion factor respectively. The $\Delta t / t_{\rm dyn}$ factor aims to explicitly satisfy the Kennicutt--Schmidt (KS) relation, which states that a fraction $f_*$ will turn into stars over a dynamical time. However, this factor is introduced at multiple points in the star formation process, which impedes the promptness of star formation and its associated feedback by only converting a limited amount of gas into stars. By opting for a timestep independent star formation, the factor $\Delta t / t_{\rm dyn}$ is removed from the calculation shown in Equation \ref{eq:mstar}, resulting in a stellar mass of
\begin{equation}\label{eq:ti_mstar}
m_{*} = m_{\mathrm{gas}} \times f_*,
\end{equation} where the symbols have the same meaning as in  Equation \ref{eq:mstar}. In this timestep independent approach, the simulation instantaneously converts $f_*$ of gas into stars in each timestep and the associated feedback will immediately start regulating further star formation. This modification greatly improves the efficiency of the star formation and feedback processes but requires further adjustments as discussed in detail in later sections. 

{\parskip 2em} As we show in Sections \ref{dwarft} and \ref{t_to_ti}, the timestep independent star formation model generally leads to a smoother buildup of stellar mass, but not without some additional effects. When we contrast the performance of the simulation, a simulation employing timestep dependent star formation will take roughly a month to complete whereas a similar setup with timestep independent star formation completes in approximately two days, reflecting the production of fewer star particles. Shorter run times allow for more exploration of the parameter space. However, this choice has significant impact on the resulting KS relation. These effects will be quantified and discussed in Section \ref{results}. In summary, we calibrate the feedback in two different star formation setups as detailed in Table \ref{tab:sfsetup}. The reasons for two setups will be discussed in Section \ref{t_to_ti}.

\begin{table}
	\caption{List of the star formation setups explored}
	\label{tab:sfsetup}
	\begin{tabular}{|p{0.1\columnwidth}||p{.22\columnwidth}|p{.22\columnwidth}|p{.22\columnwidth}|}
		\hline
		\multicolumn{4}{|c|}{Star formation setup list} \\
		\hline
		Setup & Jeans instability check & Minimum star particle mass [$M_\odot$]& Timestep dependence of star formation\\
		\hline
		1 & \cmark & $10^5$ & \cmark \\
		2 & \xmark & 0   & \xmark \\
		\hline
	\end{tabular}
\end{table}

\subsubsection{Star formation efficiency factor, $f_*$}

As mentioned, regardless of the timestep dependence of star formation, there exists an efficiency factor, $f_*$, in both Equations \ref{eq:mstar} and \ref{eq:ti_mstar}. This parameter regulates the conversion efficiency of identified gas mass in a cell into star particles: $f_*$ can vary from zero to unity but not including the limits where none or all the identified gas mass in the cell is converted to stellar mass respectively. The latter scenario will remove all the gas from the cell, resulting in a cell having a density of zero, crashing the simulation.

\subsection{Feedback parameters} \label{fbpara}

Although the creation of a star particle is immediate, feedback happens over a longer timescale, designed to mimic the gradual process of star formation. In each timestep, the star forming mass is given by  
\begin{equation}
\begin{split}
\label{eq:mform}
m_{\mathrm{form}} &= m_0\left[\left(1+\frac{t-t_0}{t_\mathrm{dyn}}\right)\mathrm{exp}\left(-\frac{t-t_0}{t_\mathrm{dyn}}\right)\right.
\\&-\left.\left(1+\frac{t+dt-t_0}{t_\mathrm{dyn}}\right)\mathrm{exp}\left(-\frac{t+dt-t_0}{t_\mathrm{dyn}}\right)\right],
\end{split}
\end{equation} 
where $m_0$ is the star particle mass, $t_0$ and $t$ are the creation time of the star particle and current time in the simulation respectively. Through this implementation, according to Equation \ref{eq:mform}, the rate of star formation increases linearly and peaks after one dynamical time before declining exponentially \citep{2011ApJ...731....6S}.

{\parskip 2em} We adopt the \citet{2011ApJ...731....6S} modification of the \citet{2006ApJ...650..560C} thermal supernova feedback model. The star particles add thermal feedback to a set of neighbouring grids with size and geometry that can be tuned by the user, known as distributed stellar feedback. This feedback continues until 12 dynamical times after its creation. In each timestep, feedback is deposited in the form of mass, energy, and metals.

{\parskip 2em} Mass is removed from the star particle and returned to the grid as gas, given by 
\begin{align} \label{eq:mej}
m_\mathrm{ej}=m_\mathrm{form} \times f_\mathrm{ej}, 
\end{align} where $f_\mathrm{ej}$ is the fraction of mass removed. The momentum of this gas is
\begin{align} 
p_\mathrm{feedback} = m_\mathrm{ej} \times v_\mathrm{particle},
\end{align} 
where $v_\mathrm{particle}$ is the velocity of the star particle and is conserved by addition into the grid cell hosting the star. The feedback energy deposited into the user defined cells is
\begin{align} \label{eq:efeedback}
E_\mathrm{feedback} = m_\mathrm{form} \times c^2 \times \epsilon, 
\end{align} where $\epsilon$ and $c$ are the feedback efficiency and speed of light respectively. For an $\epsilon$ value of $10^{-5}$ \citep{1992ApJ...399L.113C}, an energy of $\mathrm{10^{51}\,erg}$ is injected for every $\sim\ 56\,\mathrm{M_\odot}$ of stars formed. Metals are returned to the grid cells and their corresponding metallicity is given by
\begin{equation}
\begin{split}
Z_\mathrm{feedback} &= m_\mathrm{form} \times ((1-Z_\mathrm{star}) \times \eta \\
&\quad +
f_\mathrm{ej} \times Z_\mathrm{star}),
\end{split}
\end{equation}
where $Z_\mathrm{star}$ and $\eta$ are the star particle metallicity and the fraction of metals yielded from the star respectively.
 
{\parskip 2em} We assume that 25\% of the mass is removed from the star particle and returned to the grid as gas ($f_\mathrm{ej} = 0.25$) with 10\% of this returned gas being metals ($\mathrm{\eta = 0.1}$), consistent with \citet{1992ApJ...399L.113C}. These values result in a total metal yield of 0.025 of the mass of the star particle, similar to the calculations by \citet{1996MNRAS.283.1388M}. Also, this metal yield is consistent with average values in the MW, with a mean SFR of $\mathrm{\sim 3\,M_\odot\,yr^{-1}}$, a core-collapse supernova rate of 1 per 40 years, and an IMF-averaged metal yield of $\mathrm{\sim 3\,M_\odot}$ per supernova \citep{2011ApJ...731....6S}. Therefore, we leave the values of both $f_\mathrm{ej}$ and $\eta$ unaltered.

{\parskip 2em} Instead, we focus on the factors that influence the energy injection, both in terms of the amount and the physical extent. We select three factors in the feedback implementation to be varied for the calibration of the simulations. They are $\epsilon$, radius of feedback ($r$) and number of cells ($s$) within $r$. The first parameter is related to the amount of feedback energy emitted by the star particle (see Equation \ref{eq:efeedback}) while the remaining parameters work together to define the extent of energy injection. These will be described in more detail in the following sections.

\subsubsection{Feedback efficiency, $\epsilon$}

The amount of feedback energy injected as thermal energy is given by Equation \ref{eq:efeedback}. It is dependent on both rest mass energy ($m_\mathrm{form} \times c^2$) and a user-defined fraction, $\epsilon$. The former relies on the amount of stellar mass created per timestep (see Equation \ref{eq:mform}), and the latter defines the percentage of the rest mass energy injected into the IGM. Together with Equation \ref{eq:mej}, this implementation is similar to the temporal release of Galactic Superwind energy and ejected mass from stars into the IGM discussed in \citet{2006ApJ...650..560C}.

\subsubsection{Feedback energy injection extent}\label{dist}

\begin{figure}
	\begin{subfigure}[b]{0.4\columnwidth}
		\includegraphics[width=\linewidth]{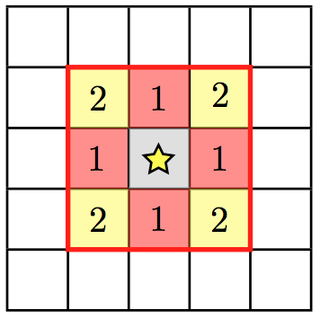}
		\label{fig:rad1}
	\end{subfigure}
	\hfill 
	\begin{subfigure}[b]{0.4\columnwidth}
		\includegraphics[width=\linewidth]{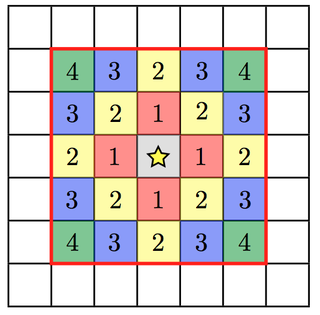}
		\label{fig:rad2}
	\end{subfigure}
\caption{Schematics of the distributed stellar feedback setup for $r=1$ (left) and $r=2$ (right) \citep{distributedfeedback}. Different coloured cells illustrate the accessible $s$ corresponding to the particular value of $r$. For example, if $r=1$, the maximum allowable value of $s$ is 3 and it corresponds to an injection of feedback energy into a $3\times3$ cube. For a detailed explanation, refer to Section \ref{dist}.}
\label{fig:dist}
\end{figure}

In the original feedback method described by \citet{2006ApJ...650..560C}, all of the feedback energy is injected into the grid cell housing the star particle. However, \citet{2011ApJ...731....6S} modified this to allow the feedback to be spread across multiple zones as a means of bypassing the overcooling issues, where too much energy injected into a single grid cell can result in unphysical short cooling times. This setup is known as distributed stellar feedback, and it is described by $r$ and $s$. These parameters work together to define the physical extent of the injection of feedback from the star particle. We can visualise it in terms of a cube surrounding the star particle in the centre. $r$ is the distance of the cell from the star particle. When $r=1$, it refers to a 3 $\times$ 3 cube since all the cells are within one cell distance away from the star particle. Similarly, when $r=2$, it refers to a $5\times 5$ cube around the star particle. These alternatives are illustrated in two dimensions in the left and right panel of Figure \ref{fig:dist} respectively. 

{\parskip 2em} The parameter $s$ gives the number of steps allowed to be taken from the star particle within the cube determined by $r$. Referring to the left panel of Figure \ref{fig:dist}, setting $s=2$ corresponds to an allowable two steps of movement away from the star particle, specifying injection within the cells labelled 1 and 2 in the $3\times 3$ cube. As the value of $r$ increases, shown in the right panel of Figure \ref{fig:dist}, so the maximum accessible value of $s$ increases. These increased values translate to more flexibility in the usage of distributed stellar feedback.

{\parskip 2em} In summary, we calibrate our simulations with $\epsilon$, $r$ and $s$, and $f_*$ to match the observations. For the remainder of the paper, when discussing the combination of parameters in a simulation setup, they will be referred to as a vector with components ($\epsilon$, $r$\_$s$, $f_s$), e.g., (1.0\e{-5}, $1\_3$, $0.1$).

\subsection{Analysis}
Haloes are identified using the Robust Overdensity Calculation using k-Space Topologically Adaptive Refinement ({\tt ROCKSTAR}) halo finder \citep{2013ApJ...762..109B}. It is a 6-dimensional phase-space finder, using both positions and velocities of particles to locate and define a halo. In regions where the density contrast is insufficient to distinguish which halo hosts a given particle, {\tt ROCKSTAR} can differentiate subhaloes and major mergers that are close to the centres of their host haloes. This feature is particularly useful in identifying main haloes when creating zoom simulations of lower mass haloes. Analysis of the simulation results is then carried out using the {\tt yt} analysis toolkit \citep{2011ApJS..192....9T}.

\begin{figure*}
	\begin{subfigure}[b]{1.0\columnwidth}
		\includegraphics[width=\linewidth]{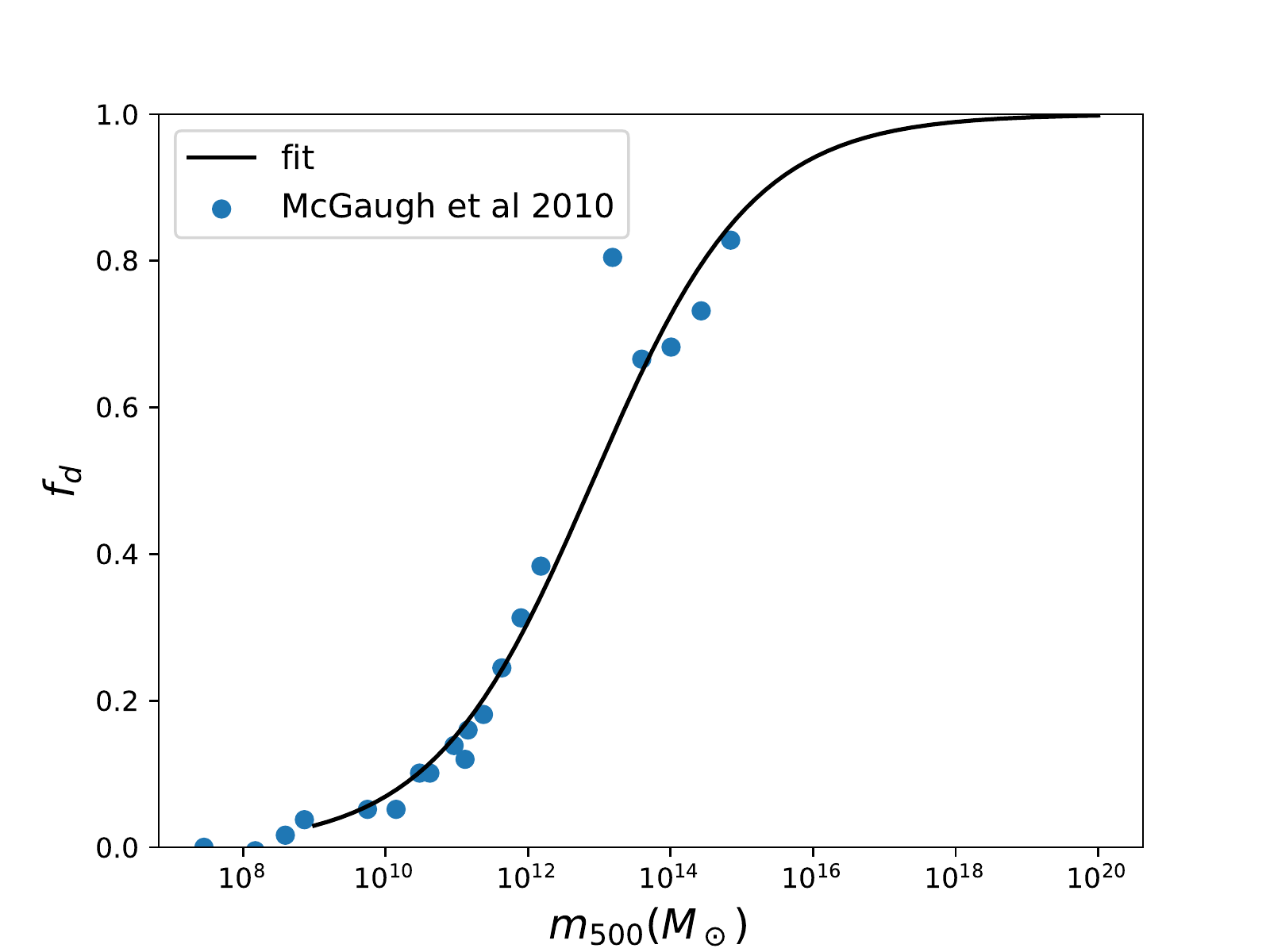}
		\label{fig:fd_fit}
	\end{subfigure}
	\hfill 
	\begin{subfigure}[b]{1.0\columnwidth}
		\includegraphics[width=\linewidth]{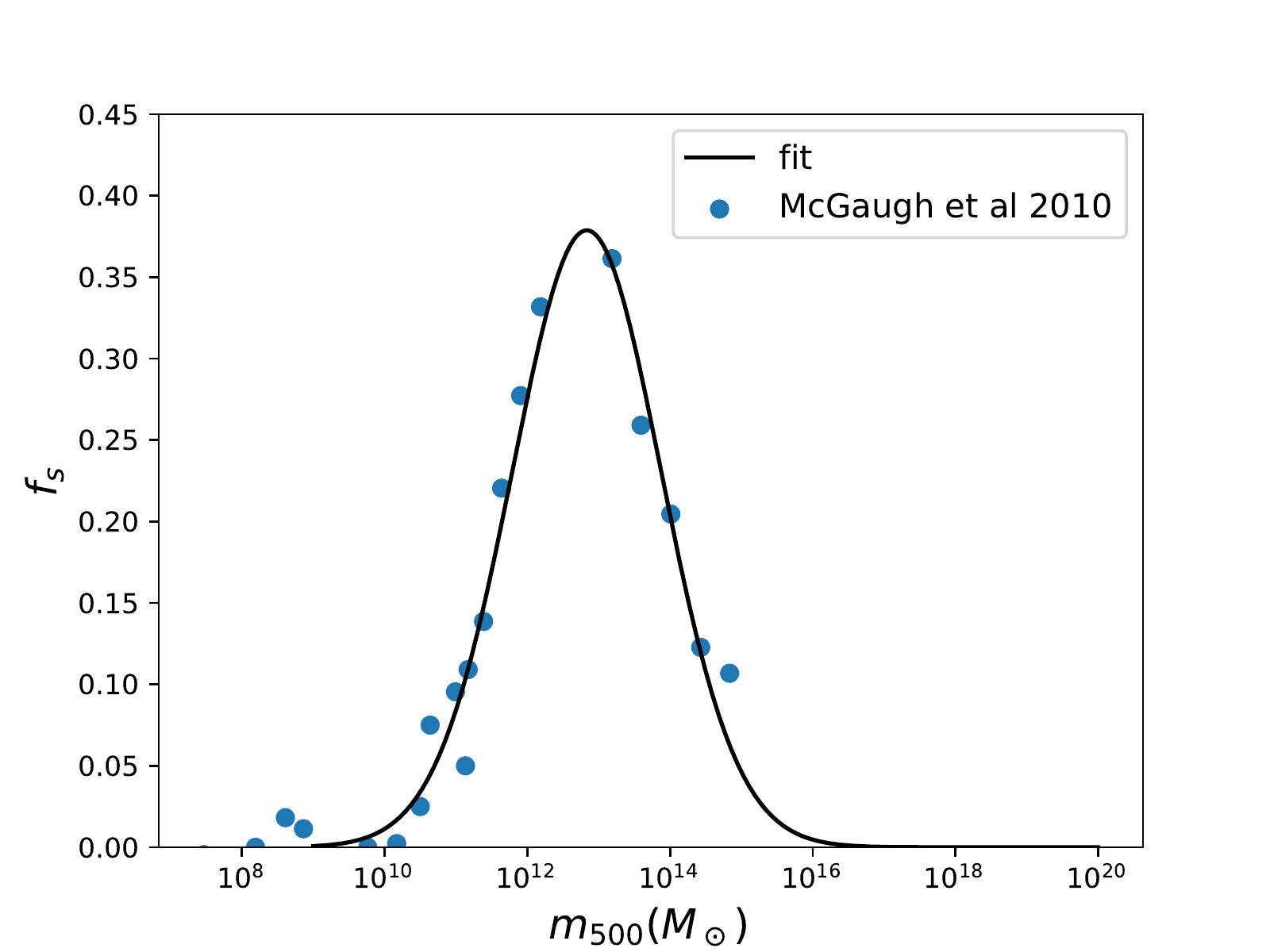}
		\label{fig:fs_fit}
	\end{subfigure}
	\caption{Best fit of the baryon fraction relative to the expected global fraction, $f_d$ (left) and the fraction of those expected baryons in the form of stars,  $f_s$ (right) from \citet{2010ApJ...708L..14M} with Equations \ref{eq:fd_fit} and \ref{eq:fs_fit} respectively.}
	\label{fig:fit}
\end{figure*}

\section{Observational Calibrations} \label{observations}

\subsection{Baryon content of cosmic structures}\label{mcgaugh}

The main observables matched in this suite of simulations are taken from the work of \citet{2010ApJ...708L..14M}, where the authors attempted to quantify the distribution of baryonic mass within cosmic structures. Galaxies are broadly categorised into rotationally supported and pressure supported systems. These are further divided into stellar dominated spiral galaxies and gas dominated galaxies for the rotationally supported system, and elliptical galaxies, local group dwarfs and some clusters of galaxies for pressure supported systems. The primary method for determining the total mass budget in the different systems is their equivalent circular velocity ($V_\mathrm{c}$) obtained through various methods described in detail in  \citet{2010ApJ...708L..14M} and summarised in Table \ref{det_met}.

\begin{table}
	\centering
	\caption{Methods of determining $V_\mathrm{c}$ for different gravitationally bound systems (see \citet{2005ApJ...632..859M, 2005ApJ...635...73H, 2007ApJ...667..176G, 2007ApJ...667L..53W, 2009ApJ...704.1274W, 2007ApJ...670..313S, 2009ApJ...703..982G}).}
	\label{det_met}
	\begin{tabular}{cc} 
		\hline
		Gravitationally bound systems & Methods \\
		\hline
		Stellar dominated spiral galaxies & Rotation velocities \\
		Gas dominated galaxies & Baryonic Tully-Fisher relation \\
		Elliptical galaxies & Gravitational lensing \\
		Local group dwarfs & Direct measurement \\
		Clusters of galaxies & Hot X-ray emitting gas \\
		\hline
	\end{tabular}
\end{table}

{\parskip 2em} In their analysis, \citet{2010ApJ...708L..14M} chose to present their results using $r_{500}$, a radius where the enclosed density is 500 times the critical density of the universe. The main result presented in Figure 2 of \citet{2010ApJ...708L..14M} relates the fraction of expected baryons that are detected,
\begin{equation}
f_d = \frac{m_b}{f_b \times m_{500}},
\label{eq:fd}
\end{equation} and the conversion efficiency of baryons into stars, 
\begin{equation}
f_s = \frac{m_*}{f_b \times m_{500}},
\label{eq:fs}
\end{equation} where $m_b$ and $m_{500}$ refer to the baryonic and total mass within this radius respectively, and $f_b$ is the universal baryon fraction determined to be $0.17 \pm 0.01$ \citep{2009ApJS..180..330K}. One important point to note is that these fractions are dependent on the choice of radius. To facilitate comparison of our results with this paper, we produced the following fitting formula to the data from Figure 2 in \citet{2010ApJ...708L..14M}, which we illustrate in Figure \ref{fig:fit}:
\begin{equation}
f_d = \frac{1}{1+e^{-x}},
\label{eq:fd_fit}
\end{equation} and
\begin{equation}
f_s = 0.91\exp\left(\frac{-y^2}{2}\right) \times f_d,
\label{eq:fs_fit}
\end{equation} where 
\begin{equation}
x = \frac{{\log_{10}}\left(m_{500}/M_\odot\right) - 12.91}{1.12},
\end{equation} and
\begin{equation}
y = \frac{{\log_{10}}\left(m_{500}/M_\odot\right) - 12.19}{1.18}.
\end{equation} We aim to calibrate our suite of simulations to yield a good match to these fits. Also, we will compare our simulated galaxy properties to the Kennicutt--Schmidt relation, which serves as an additional constraint.

\subsection{Kennicutt--Schmidt relation} \label{ks-relation}

The KS relation is a measure of the correlation between gas surface density and the SFR per unit area. From the work of \citet{1959ApJ...129..243S, 1989ApJ...344..685K, 1998ApJ...498..541K, 2007ApJ...671..333K,2008AJ....136.2846B}, there appears to be a tight correlation between these measured properties on galactic scales ($\sim$ kpc). This strong relation makes it one of the critical observations that simulations with star formation attempt to match.

{\parskip 2em} We adopt a similar methodology to that of the AGORA project \citep{2016ApJ...833..202K}. The SFRs are calculated using the mass of star particles and time-averaged over the past $20\,{\rm Myrs}$ of the simulation snapshot. Together with the gas density they are then deposited onto a fixed resolution grid of $750\,{\rm pc}$, consistent with the methodology of \citet{2008AJ....136.2846B}, to derive the SFR and gas surface density required by the KS relation. In fact, we find that the conclusions drawn are insensitive to changes in the grid resolution. With non-zero SFR surface density patches, we will also compare our results to 
\begin{equation}
\mathrm{\log\Sigma_{SFR} = 1.37 \log\Sigma_{gas}-3.78},
\label{eq:ks_fit}
\end{equation} which is obtained from the best observational fit given by Equation 8 in \citet{2007ApJ...671..333K}. 

\section{Results}\label{results}

\subsection{MW galaxy zoom simulations with Setup 1}\label{mwt}

\begin{figure*}
	\includegraphics[width=\linewidth]{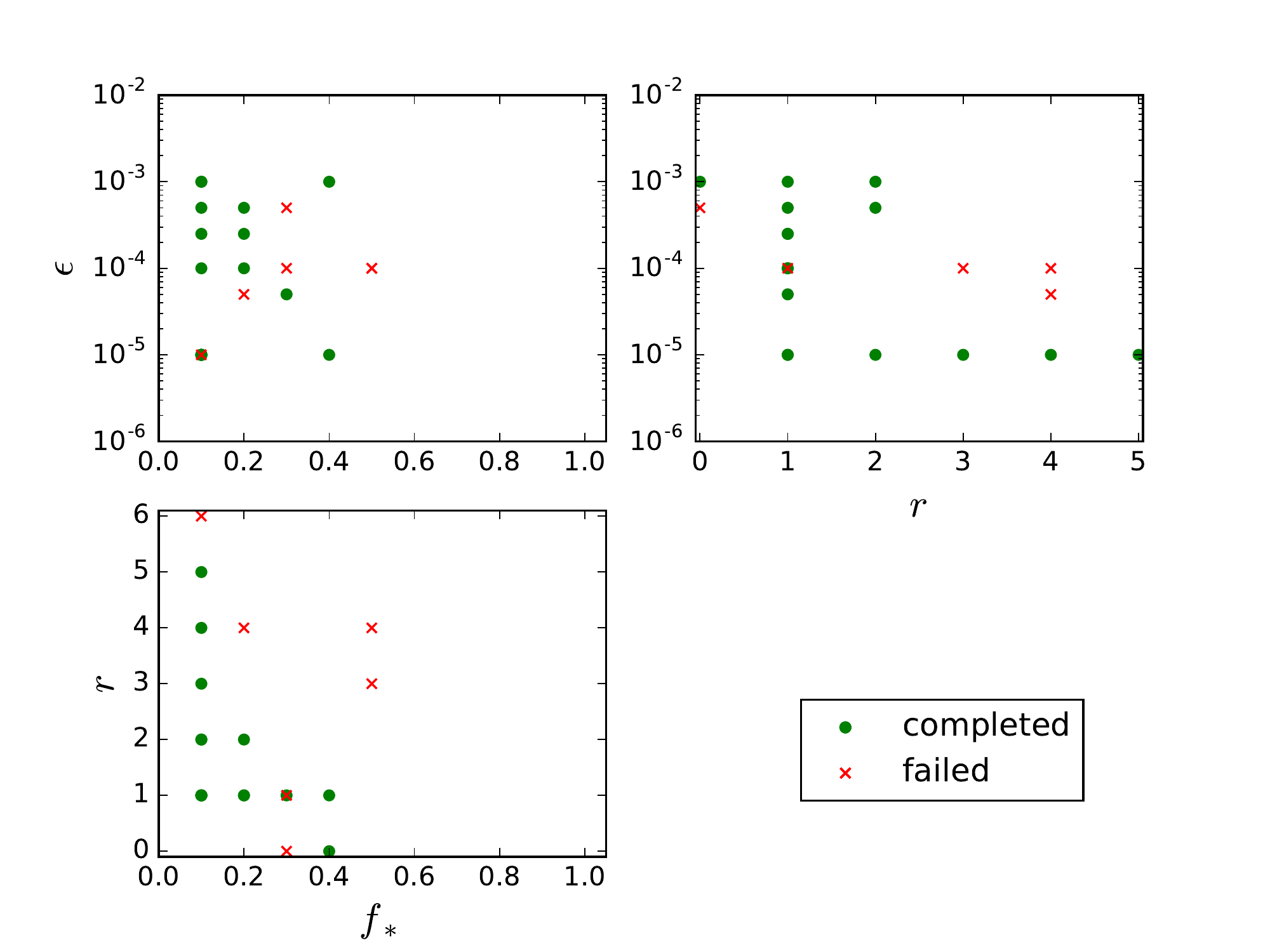}
	\caption{Overview of parameter space exploration using a total of 22 different combinations of $f_*$, $\epsilon$ and $r$ with Setup 1. The plots are $\epsilon$ against $f_*$ (top left), $\epsilon$ against $r$ (top right) and $r$ against $f_*$ (bottom left). The green dots and red crosses represent runs that reached and failed to reach $z=0$ respectively. We can identify regions of parameter space more likely to result in the inability of the simulation to reach $z=0$ and the causes are explained in more detail in Section \ref{mwt}.}
	\label{fig:t_para}
\end{figure*}

We explore the parameter space by switching on the Jeans instability check, applying timestep dependent star formation and setting a threshold star particle mass of $10^5\,\mathrm{M_\odot}$ (Setup 1); see Table \ref{tab:sfsetup}. With this setup, we run a total of 22 simulations by modifying $f_*$ (see Equation \ref{eq:mstar}), $\epsilon$ (see Equation \ref{eq:efeedback}), $r$ and $s$ (see Section \ref{dist}), as shown in Figure \ref{fig:t_para}. This explored region of parameter space is motivated both physically and numerically.  \citet{1992ApJ...399L.113C} applied a value of $\epsilon$ ($10^{-4.5}$), which is similar to other work \citep{1981ApJ...243L.127O, 1987Natur.326..455D}. The values of $r$ and $s$ are restricted by the maximum number of cells used to define a grid. Lastly, we can constrain the range of values that $f_*$ can take with the ratio of $f_s$ to $f_d$. From \citet{2010ApJ...708L..14M}, $f_*$ is limited between 0.1 and 0.9 approximately across the halo mass range.

{\parskip 2em} Out of the 22 simulations, we classify the runs into those that reached $z=0$ (completed) and those that did not (failed) since we are interested in the relevant properties at $z=0$. A fraction of the simulations were unable to reach the final redshift due to unrecoverable errors in the hydrodynamics solver, mostly associated with extreme star formation and/or feedback parameters. Since the failed simulation contains extreme feedback parameters, e.g, large amount of feedback energy, it is unlikely that this prescription will result in the best match to the observed properties, presented in Section \ref{mcgaugh_time_sim}. Overlapping points with conflicting conclusions exist in Figure \ref{fig:t_para} because as we are showing the 2-dimensional projection of the 3-dimensional parameter space.

\begin{figure*}
    \vglue-3em
	\includegraphics[width=0.76\linewidth]{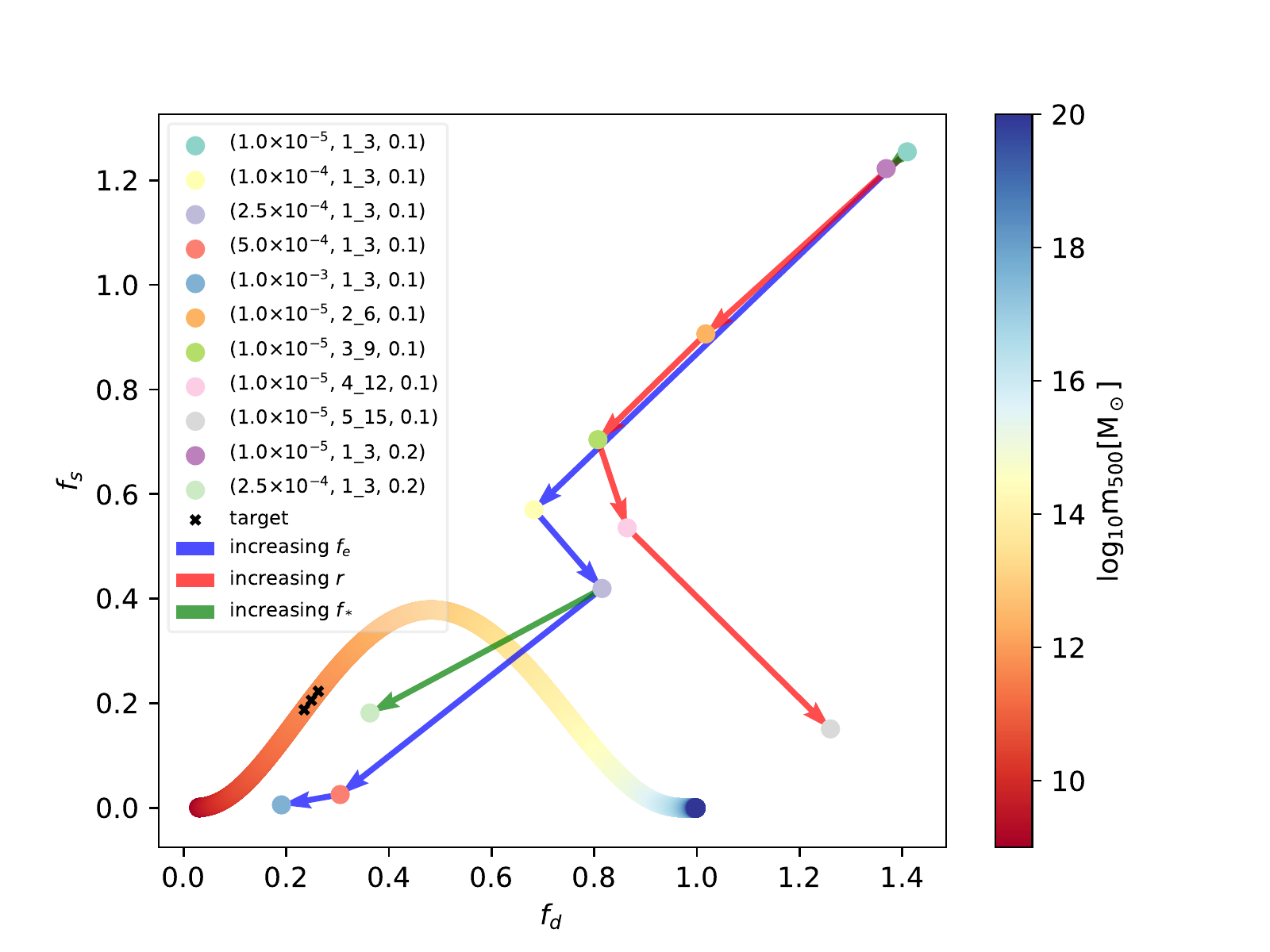}
	\caption{Graph of the stellar mass fraction $f_\mathrm{s}$ against the detected baryon fraction $f_\mathrm{d}$ across a range of $m_{500}$. The colour coded curve shows the values of  $f_\mathrm{s}$ and $f_\mathrm{d}$ corresponding to the mass range indicated. Cross marks on this curve shows the values to be matched. It includes the mean of $m_{500}$ with the upper and lower limits given by the maximum and minimum $m_{500}$ from the simulations plotted. Dots represent simulation with different feedback parameters and contrasting coloured arrows show effects of increasing a particular parameter. From the ensemble of simulations, (2.5\e{-4}, 1\_3, 0.2) yielded the most realistic baryonic makeup for the simulated MW galaxy. For a detailed description, refer to Section \ref{mcgaugh_time_sim}. }
	\label{fig:trend}
\end{figure*}

\begin{figure}
\begin{center}
	\strut\includegraphics[width=0.51\textwidth]{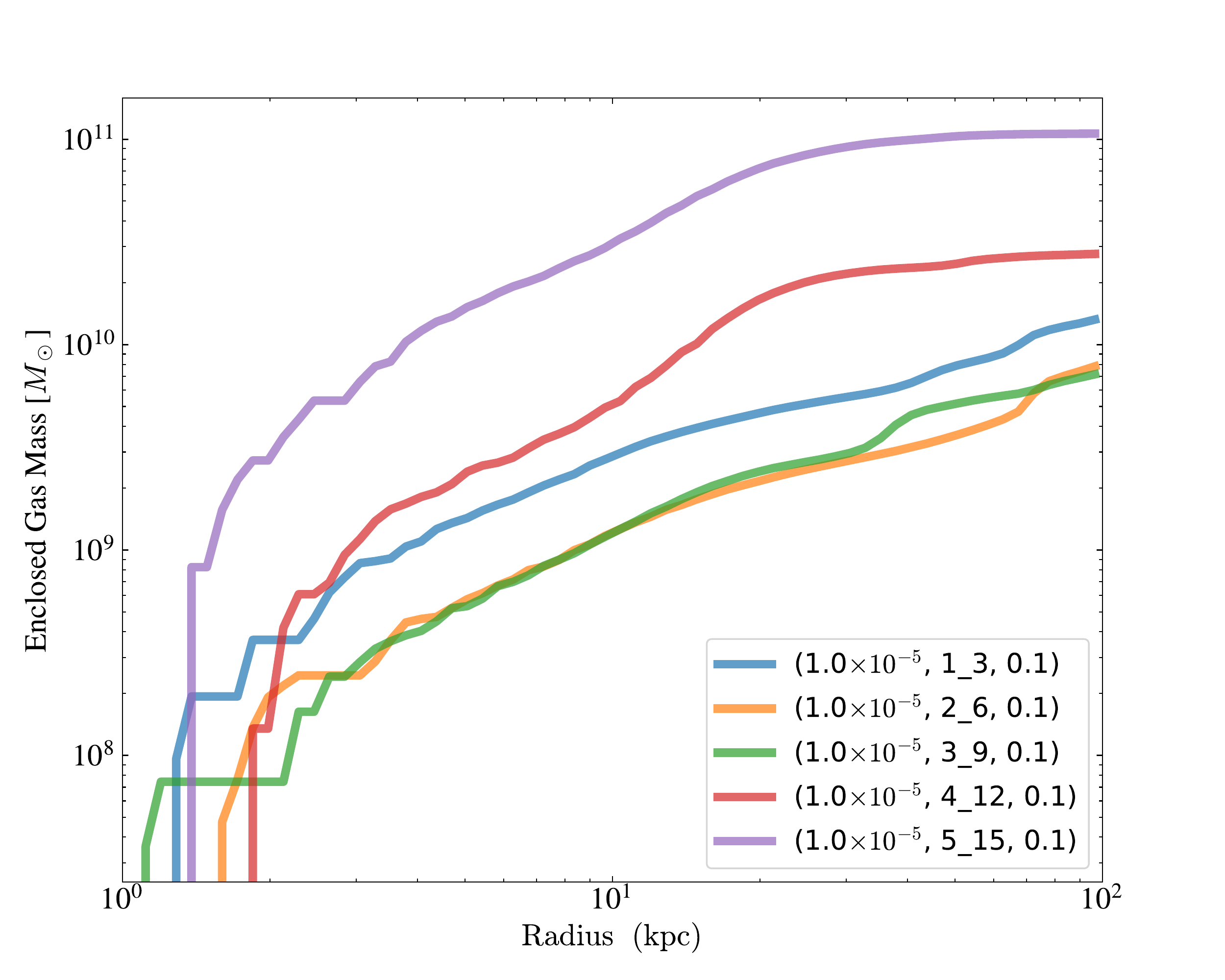}
	\caption{Cumulative plot of $m_\mathrm{gas}$ against halo radius. The different coloured lines represent simulations of various $r$ and $s$. As the extent of feedback injection is increased beyond $r$ = 3 and $s$ = 9, the amount of baryons in the outer region of the halo is significantly higher. This trend highlights the inhibition of gas collapse to form stars as the extent increases, providing support for the explanation given in Section \ref{mcgaugh_time_sim} for the trend in Figure \ref{fig:trend}.} 
	\label{fig:cell_mass}
	\end{center}
\end{figure}

\subsubsection{Comparison to baryonic properties from \citet{2010ApJ...708L..14M} -- Setup 1}\label{mcgaugh_time_sim}

Initially, we attempted to cover the parameter space optimally with minimal numbers of simulations using Latin Hypercube Sampling \citep{10.2307/1268522}. We wanted to minimise the maximal distance between various points in our feedback parameter space as described by \citet{2009ApJ...705..156H}. However, due to the failure of several runs to reach $z=0$, it is not possible to obtain a space-filling design. Therefore, we try a more fundamental approach to quantify how changing each parameter will affect the observables. This result is presented in Figure \ref{fig:trend}, showing a plot of $f_\mathrm{s}$ against $f_\mathrm{d}$ across a range of $m_{500}$. 

{\parskip 2em} From the initial values of (1.0\e{-5}, 1\_3, 0.1), we vary $\epsilon$ only, which corresponds to a change in the strength of feedback. Increasing the strength of feedback reduces both the $f_s$ and $f_d$ parameters of the halo (see the blue arrow in Figure \ref{fig:trend}). This evolution can be easily explained by the increased expulsion of gas due to stronger feedback, reducing the amount of fuel available to form stars, which leads to a decrease in $f_s$. The removal of gas also causes the amount of baryons within $r_{500}$ or $f_d$ to decrease. 

{\parskip 2em} We then try to increase $f_*$. This change has a direct impact on the total stellar mass as more gas mass is converted into stars. However, this increased star formation yields stronger feedback. Therefore, the net result of increasing $f_*$ is similar to increasing $\epsilon$, which decreases both $f_s$ and $f_d$ (see the green arrow in Figure \ref{fig:trend}). To a lesser extent, however, this effect is evident from the small transition of the cyan point to the purple point on the top right of the plot. To improve the clarity of an increase in $f_*$, we add another green arrow connecting another set of data points (grey and light green dots). This difference in the impact of $f_*$ also suggests its sensitivity to other feedback parameters.   

{\parskip 2em} The last parameters to adjust are $r$ and $s$. Essentially, we are increasing the size of the cube into which the feedback energy is injected (see Figure \ref{fig:dist}). By increasing $r$ (and, correspondingly, $s$), $f_s$ and $f_d$ are reduced, similarly to the effect of increasing $\epsilon$ and $f_s$. However, this phenomenon only persists until $r = 3$ and $s = 9$, which corresponds to a $7^3$ box or 343 cells centred around the star particle. Beyond this point, the trend changes when a further extension of the feedback injection decreases $f_s$ but increases $f_d$, indicating the presence of a turnaround point. As energy is deposited further from the star particle, the gas is kept away at a larger distance from the centre of the gravitational potential well as seen in Figure \ref{fig:cell_mass}. As a result, $m_*$ decreases as fewer stars form due to a deprivation of fuel for star formation while $m_\mathrm{gas}$ increases as more gas is now present. Increasing the physical extent of feedback injection beyond $r = 3$ and $s = 9$ only serves to dilute the amount of feedback energy per cell, leading to gas remaining near the virial radius of the halo. Thus $f_d$ increases while $f_s$ decreases. Furthermore, the average number of cells within a single grid in an {\tt Enzo} simulation is not likely to be much larger than about $7^3$, so extending beyond $r = 3$ and $s = 9$ should be avoided as feedback is only deposited on the local grid.

{\parskip 2em} From the 16 completed simulations, the combination of parameters that yielded the most MW-like properties in the halo is (2.5\e{-4}, 1\_3, 0.2), which is represented by the pale green dot in Figure \ref{fig:trend}. The halo contains a stellar mass comparable to the MW while having approximately 50\% more baryon mass than the MW halo. This point is the closest match to the target for the region of parameter space that we sampled. The next best set of parameters that produce halo properties matching the target is (5.0\e{-4}, 1\_3, 0.1). While it provides a better $f_d$ agreement, the value of $f_s$ is approximately zero. From the trends and the best match in Figure \ref{fig:trend}, further improvement in the agreement of halo properties will only be marginal. In order to achieve a better agreement, we suggest including other free parameters or even modifying the star formation and feedback model. Furthermore, this set of parameter is determined for a quiescent halo as discussed earlier. Success of this calibrated feedback prescription is likely to be dependent on the growth history as well. However, it is not within the scope of this work to design such modifications or test the robustness of our calibration against different merger histories.

\begin{figure*}
	\includegraphics{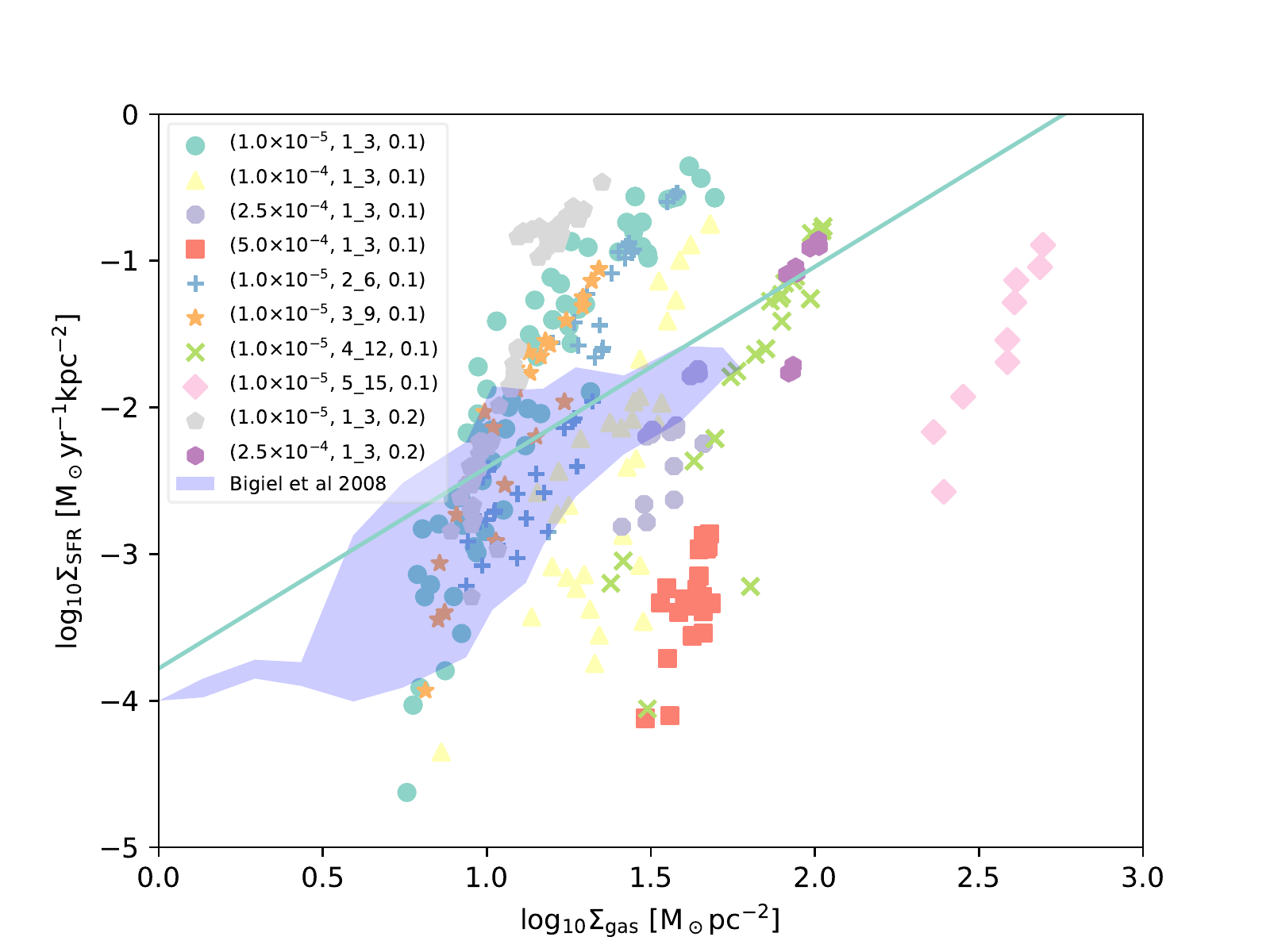}
	\caption{A graph of SFR surface density against gas surface density illustrating the KS relation. Different coloured points are simulation data from sub-kpc resolution consistent with rough approximation from the observations in nearby galaxies by \citet{2008AJ....136.2846B} represented by the blue hatched contours. The blue line is derived from the observational fit of \citet{2007ApJ...671..333K}. There is overlap between the simulation and observation but there are differences that will be discussed in Section \ref{kst}.}
	\label{fig:ks_t}
\end{figure*}

\subsubsection{Kennicutt--Schmidt relation -- Setup 1} \label{kst}

As discussed earlier, the KS relation provides an additional constraint on the feedback calibration beyond the global baryon makeup of a MW halo.To apply this constraint, we use the methodology described in Section \ref{ks-relation} to compare and contrast with the observed KS relation (blue line) in Figure \ref{fig:ks_t}. We also include a rough approximation of the observed values of nearby galaxies from \citet{2008AJ....136.2846B} in the form of blue hatched contours.

{\parskip 2em} The results show that the simulation data intersect with observations and the fit given by Equation \ref{eq:ks_fit}, but the slopes of the simulation data differ from the KS relation in every feedback prescription. Most of our simulations manage to reproduce the characteristic `threshold' gas density value of approximately $10\,\mathrm{M_\odot pc^{-2}}$, which marks the transition point between high and low star formation efficiency and is apparent from the blue hatched contour. The slopes of the relations in the simulations do not appear to be significantly different from each other despite changes in the subgrid physics parameters. However, they are consistently steeper than the gradient of the observed relation. 

{\parskip 2em} When increasing $\epsilon$, we observe a shift towards lower SFR but higher gas density from the transition between the green circles (1.0\e{-5}, 1\_3, 0.1) and the red squares (5.0\e{-4}, 1\_3, 0.1). This shift can be explained by the higher feedback energy budget associated with a larger $\epsilon$ value, which inhibits further star formation. The simulation data points are insensitive to any increase in $r$ until $r=3$. Beyond which, comparable SFR densities are associated with higher gas densities (compare green crosses ($r=4$) and pink diamonds ($r=5$)). This trend is consistent with the explanation provided for Figure \ref{fig:cell_mass}. Lastly, from the data points of (1.0\e{-5}, 1\_3, 0.1) and (1.0\e{-5}, 1\_3, 0.2), it appears that increasing $f_*$ does not affect the relation significantly.

{\parskip 2em} The best parameter values (purple hexagon) lie along with the KS relation fit but deviate from observations as they are clustered around high gas densities. This discrepancy with \citet{2008AJ....136.2846B} suggests that this combination of $\epsilon$ and $f_*$ is too weak to create patches of lower gas surface density. However, adjustment of either factor will, in turn, affect $f_s$ and $f_d$, leading to a halo that reproduces the KS relation instead of the observations of \citet{2010ApJ...708L..14M}.

\subsubsection{Haloes in the high-resolution region -- Setup 1}\label{sat_t}

\begin{figure*}	
	\begin{subfigure}[b]{1.0\columnwidth}
		\includegraphics[width=\linewidth]{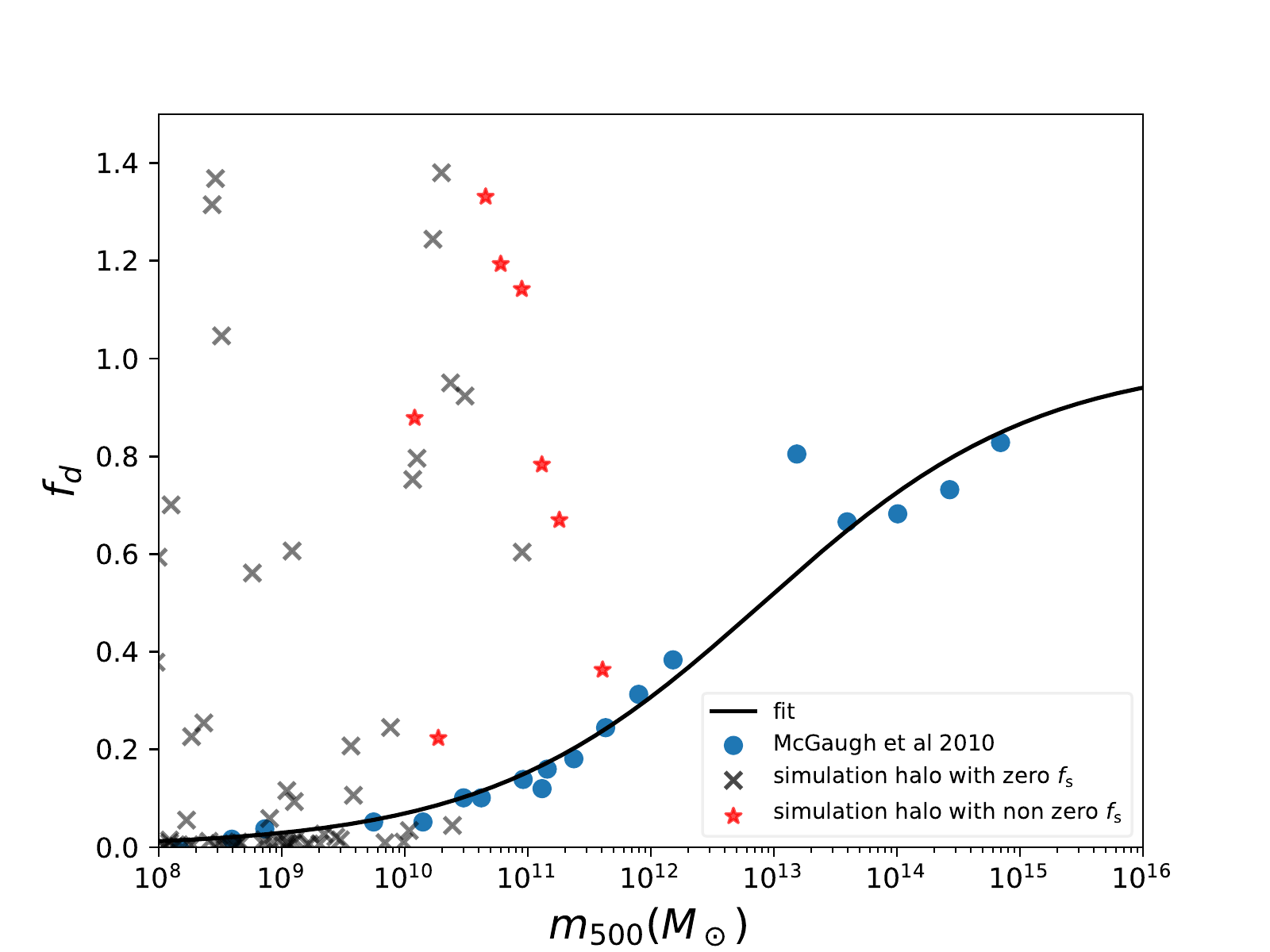}
		\label{fig:sat_fd}
	\end{subfigure}
	\hfill 
	\begin{subfigure}[b]{1.0\columnwidth}
		\includegraphics[width=\linewidth]{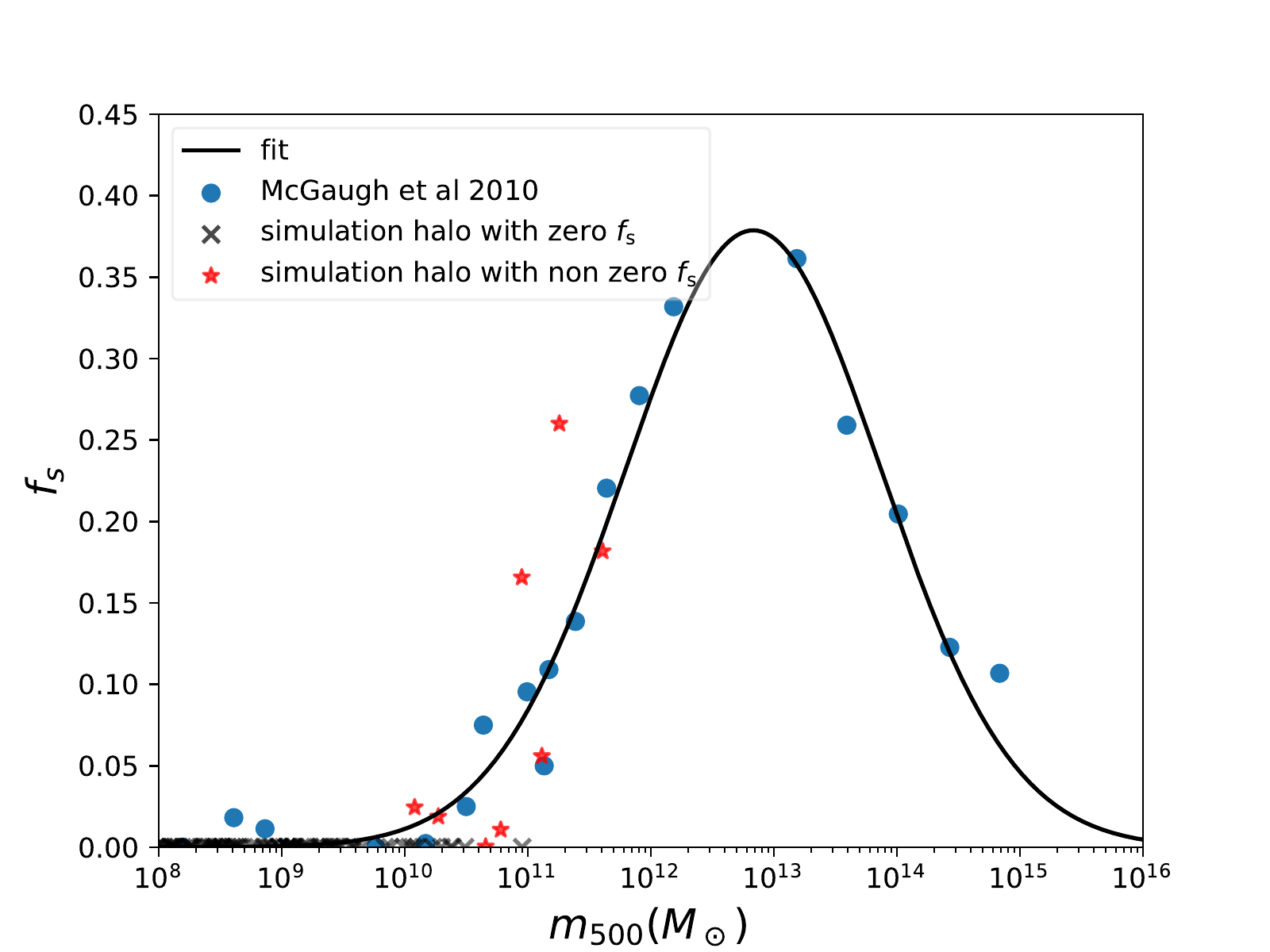}
		\label{fig:sat_fs}
	\end{subfigure}
	\caption{Graph of (a) the detected baryon fraction $f_\mathrm{d}$ against $m_{500}$ and (b) the stellar mass fraction $f_\mathrm{s}$ against $m_{500}$ with equations described in Section \ref{mcgaugh}. The black line represents the fit given by Equation \ref{eq:fd_fit} and \ref{eq:fs_fit}. The blue dots are points from \citet{2010ApJ...708L..14M} and the crosses and stars are properties of haloes with various mass from the must refined region in the simulation. The black cross and red star refer to haloes in the must refine region with zero and non zero $f_\mathrm{s}$ respectively. Other than the most massive halo (MW halo), other haloes struggle to contain the appropriate amount of stars and gas. Refer to Section \ref{sat_t} for a detailed description.}
	\label{fig:mcgaugh_t_sat}
\end{figure*}

Since we specify a safety factor of three virial radii to prevent contamination of the MW halo in the zoom simulation, there are other central and satellite haloes of varying mass in this region. Figure \ref{fig:mcgaugh_t_sat} illustrates the properties of other central haloes in the simulation with the best feedback prescription of (2.5\e{-4}, 1\_3, 0.2). This plot is not presented in a similar way to Figure \ref{fig:trend} because we are looking at a range of halo masses. Instead, we populate Figure \ref{fig:fit} with the corresponding $f_s$ and $f_d$ of various central haloes in the high-resolution region of the MW galaxy zoom simulation.  

{\parskip 2em} We present the graph of $f_\mathrm{d}$ against $m_{500}$ on the upper panel and $f_\mathrm{s}$ against $m_{500}$ on the lower panel of Figure \ref{fig:mcgaugh_t_sat}. The black lines are Equations \ref{eq:fd_fit} and \ref{eq:fs_fit} fitted to the observations (blue dots). From the graph of $f_\mathrm{s}$ on the lower panel, the properties either match observation well or do not form stars at all ($f_\mathrm{s} = 0$), which is represented by a red star and grey cross respectively. The same cannot be said for the $f_\mathrm{d}$ relation on the left where there are vastly different properties with no apparent relation to the $f_\mathrm{d}$ values. We see lower mass haloes with $f_\mathrm{d}$ reaching 1.4, exceeding that of the universal baryon fraction while possessing $f_\mathrm{s}$ that is close to observations. These haloes are in contrast to the MW halo (rightmost red star), which hints at the need for additional modifications required to understand and determine if this discrepancy is a numerical byproduct due to the fractional mass resolution of the lower mass halo. Therefore, we attempt a zoom simulation of a dwarf galaxy around $10^{10}\,\mathrm{M_\odot}$ with a comparable mass resolution to the MW zoom to investigate if the conclusion from Figure \ref{fig:mcgaugh_t_sat} is due to resolution and whether this feedback prescription is universal.

\subsection{Dwarf galaxy zoom simulations with Setup 1}\label{dwarft}
Using the combination of parameters (2.5\e{-4}, 1\_3, 0.2), we implement the feedback prescription in a dwarf galaxy with a mass of approximately $10^{10}\,\mathrm{M_\odot}$. However, the results indicate an absence of stars within the halo, consistent with Figure \ref{fig:mcgaugh_t_sat}. Reviewing the star formation routine (see Section \ref{sfmod}), we find that the Jeans instability check is the bottleneck of star formation. Due to the spatial resolution implemented in this dwarf galaxy, according to the discussion in Section \ref{usejeansmass}, the Jeans instability check restricts star formation that should occur in reality. Therefore, to allow star formation, we switch off this Jeans instability check in the star formation routine. We label such runs as NJ.

\begin{figure}
	\includegraphics[width=\linewidth]{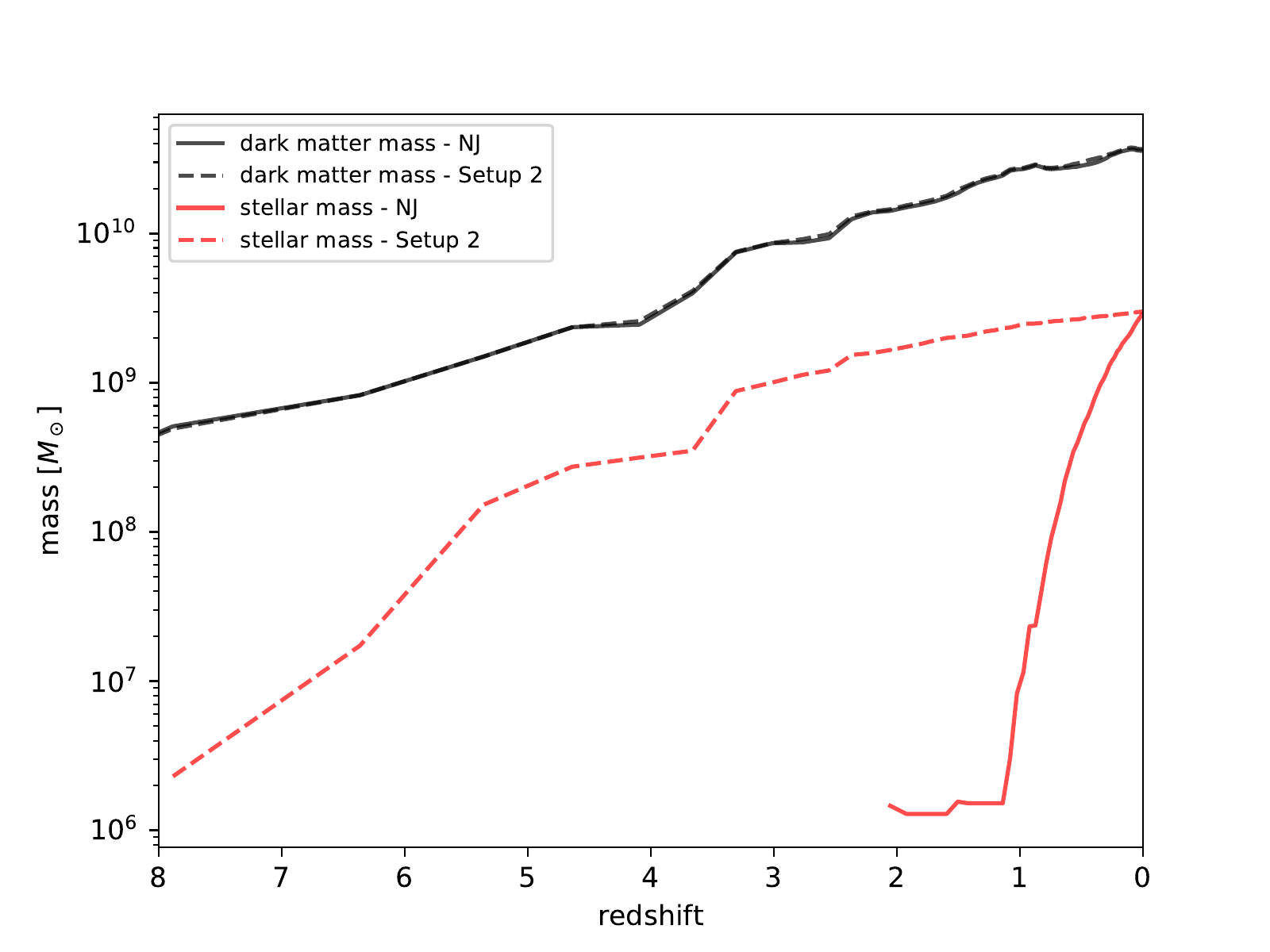}
	\caption{Redshift evolution of dark matter and stellar mass in the dwarf galaxy. Solid and dashed lines are mass evolution of NJ runs and Setup 2 runs respectively. Black lines refer to the dark matter mass evolution while red lines refer to the stellar mass evolution in the halo. By only switching off the Jeans instability check, the dwarf galaxy starts to form stars, albeit only at $z\approx 2$, which is too late. Therefore, we need a full transition to Setup 2 where the minimum star particle mass is set to zero in order to allow for star formation at an earlier time. Refer to Section \ref{dwarft} for discussion.} 
	\label{fig:dwarf_mass_evo}
\end{figure}

{\parskip 2em} Figure \ref{fig:dwarf_mass_evo} illustrates the virial (black) and stellar mass evolution (red) in the dwarf galaxy with different setups (solid vs dashed lines). As expected, removing the Jeans mass criterion allows stars to form in the dwarf galaxy zoom simulation (solid red line). However, star formation starts around $z=2$, which is late as compared to the MW zoom simulation, for which star formation commenced at $z\simeq6.5$. Further investigations yielded the conclusion that the star formation threshold mass is the next limiting factor. Therefore, we reduce the threshold mass for star particle creation to zero, which relaxes the condition for star formation, allowing star particles to be created at $z=8$ in the simulation. On top of these changes, we switch off the timestep dependence of star formation. This results in Setup 2 as shown in Table \ref{tab:sfsetup}.

{\parskip 2em} The purpose of $\Delta t/t_\mathrm{dyn}$ in Equation \ref{eq:mstar} is to ensure the adherence of star formation to the KS relation. However, in Equation \ref{eq:mform} where feedback is modelled to occur across time, there are additional factors of $\Delta t/t_\mathrm{dyn}$ present to regulate these processes according to the KS relation. Hence, by switching to timestep independent star formation, we improve the promptness of the feedback. Lastly, since star formation is now instantaneous once conditions are met, high-density regions of gas are absent, reducing the time used to calculate the hydrodynamic evolution in the simulation. This absence of high-density gas is evident from the number of timesteps required for the evolution to reach $z=0$ and the time per timestep. For an identical feedback prescription, Setup 1 takes 1263 timesteps and $\sim 435s$ per timestep to reach $z=0$, which is in stark contrast to Setup 2 where it takes 663 timesteps and $\sim 125s$ per timestep for the simulation to reach $z=0$. The net result is an improvement in the speed of completion of simulations from weeks to days. 

{\parskip 2em} In summary, we modify the setup to switch off the Jeans instability check, turn off the timestep dependence of star formation and remove the requirement of a minimum star particle mass. This results in Setup 2 shown in Table \ref{tab:sfsetup}. This setup enables us to recover a more realistic star formation history beginning at $z\sim8$ (see Figure \ref{fig:dwarf_mass_evo}), which is the main motivation for the switch in setup. However, we do not compare the properties of the dwarf galaxy to observations for reasons that will be explained in Section \ref{t_to_ti}.

\subsection{Simulations with Setup 2}\label{t_to_ti}

\begin{figure}
	\includegraphics[width=\linewidth]{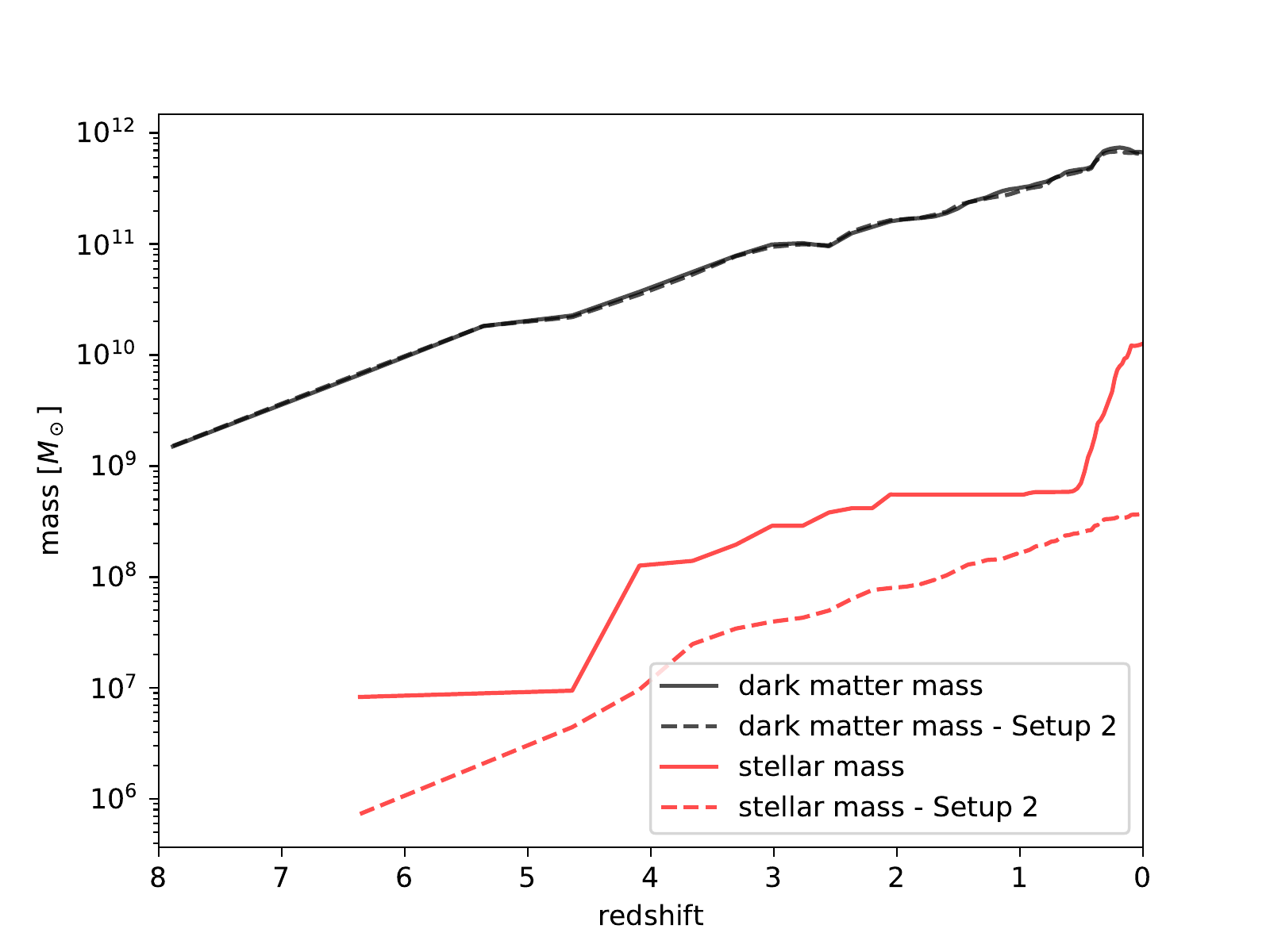}
 	\caption{Redshift evolution of dark matter and stellar mass in the MW halo. Solid and dashed lines are mass evolution of NJ runs and runs with Setup 2 respectively. Black lines refer to the dark matter mass evolution while red lines looks at the stellar mass evolution in the halo. By changing the star formation conditions, we obtain a smoother stellar mass evolution while not affecting the dark matter mass evolution. Refer to Section \ref{t_to_ti} for discussion.} 
 	\label{fig:mw_mass_evo}
\end{figure}

Due to star formation issues in the dwarf galaxy zoom simulations, we make significant changes in the simulation setup. In Section \ref{dwarft}, we show that the stellar mass of a dwarf galaxy at $z=0$ changed from zero to $\sim10^9\,\mathrm{M_\odot}$ by switching to Setup 2. We now have to review the results of the MW galaxy presented in Section \ref{mwt}. Figure \ref{fig:mw_mass_evo} shows the evolution of the dark matter and stellar mass of the MW halo in different setups. The lines and labels are similar to Figure \ref{fig:dwarf_mass_evo}.

{\parskip 2em} From the identical dark matter mass evolution in Figure \ref{fig:mw_mass_evo} for different setups (black lines), we know that we are comparing the same halo across simulations. However, the stellar mass evolution paints a different picture. Comparing both setups, although the haloes start forming stars at the same time ($z\simeq6.5$), the simulation using Setup 2 has a lower initial and final stellar mass as a result of its corresponding relaxed star formation conditions. With the minimum mass of the star particles set to zero, the stars are allowed to form with a smaller mass, which explains a lower starting point in Setup 2. Between $z = 1$ and $z = 0$ in Setup 1, we note a spike in stellar mass due to the build up of gas eligible for star formation (see Section \ref{timesf}). Despite these differences in the star formation history, the most significant one is the stellar mass of the halo at $z = 0$. The final stellar mass of the MW halo in Setup 1 is approximately $10^{10}\,\mathrm{M_\odot}$, which is two orders of magnitude higher than that in the new run with a value of roughly $10^{8}\,\mathrm{M_\odot}$. This difference means that these haloes have vastly different $f_\mathrm{d}$ and $f_\mathrm{s}$. 

{\parskip 2em} Due to the non-linear coupling of the various processes, changing individual prescriptions always requires new parameter fitting \citep{2015MNRAS.450.1937C}. With a new star formation setup, we have to re-explore the feedback parameter space with Setup 2. However, we have two distinct advantages as compared to before. The first is that we understand the general effects changing the feedback parameters have on the $f_s$ and $f_d$ of the halo (see Figure \ref{fig:trend}). Secondly, the simulations will complete much faster, allowing us to obtain more data points, both in general feedback parameter space and in the region around the best match to observations. This improvement will help us narrow down the feedback prescription, and possibly identify more than one combination that yields a close match. Obtaining more than one set of parameters will open up the possibilities of testing the robustness of the feedback prescription in the MW halo zoom simulations, haloes in the high-resolution region and the dwarf galaxy zoom simulations.

\begin{figure*}
	\includegraphics[width=0.80\linewidth]{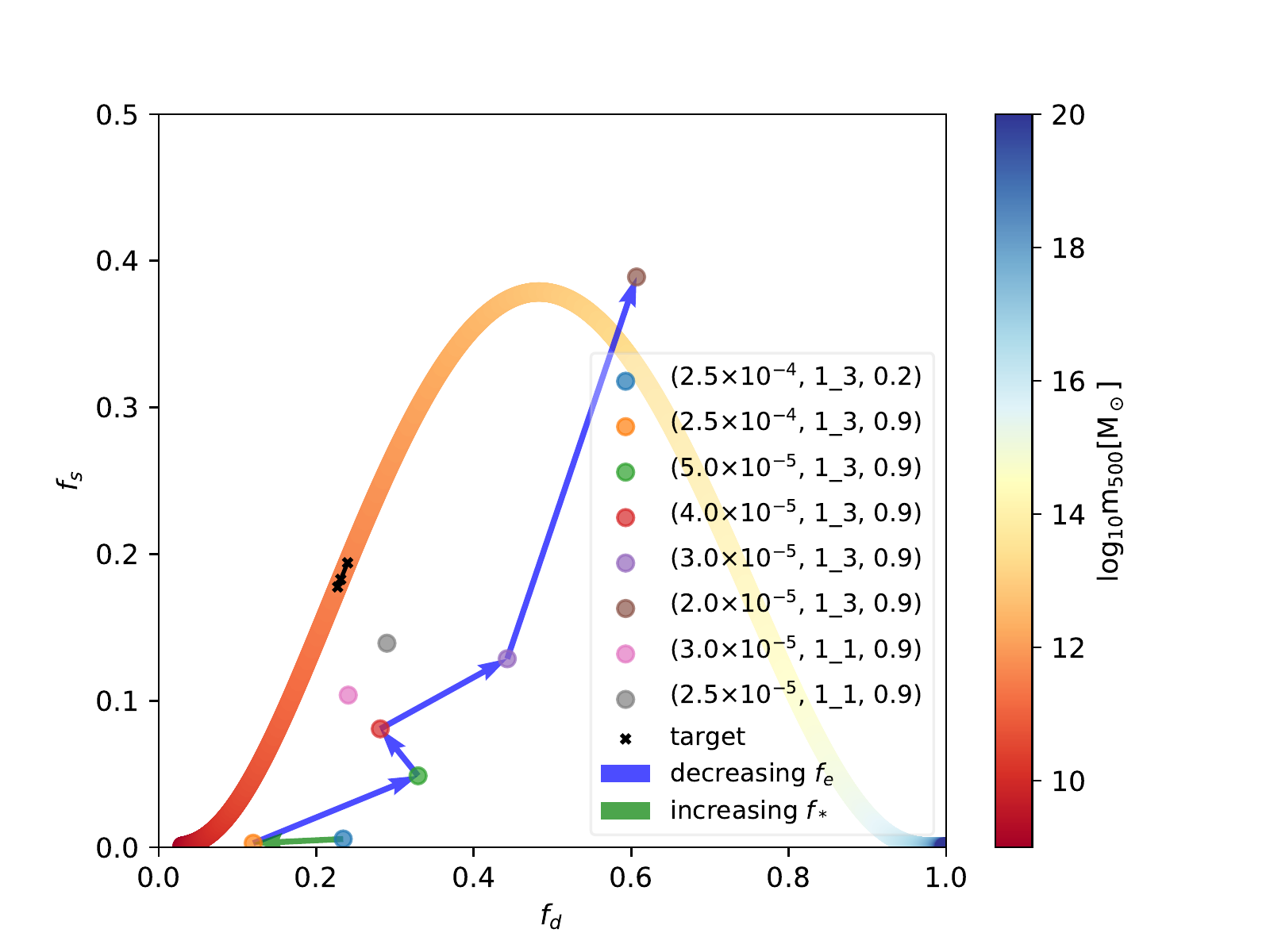}
	\caption{Graph of $f_\mathrm{s}$ against $f_\mathrm{d}$ across a range of $m_{500}$ for Setup 2. The symbols and colour bar have the same meaning as those shown in Figure \ref{fig:trend}. The best combination of feedback parameter (blue dot) in Setup 1 no longer produces a realistic baryonic makeup of the MW halo. Instead, we have to re-calibrate the star formation and feedback prescription with the trends from Figure \ref{fig:trend}, resulting in (2.5\e{-5}, 1\_1, 0.9) and (3.0\e{-5}, 1\_1, 0.9) as the values required for Setup 2. For detailed description, we refer to Section \ref{mcgaugh_ti_sim}.}
	\label{fig:ti_trend}
\end{figure*}

\begin{table*}
	\caption{List of feedback prescriptions discussed in Sections \ref{mwt}, \ref{mwti}, \ref{dwarfti} and \ref{chaos} with the relevant properties of the halo of interest. These include $m_{500}$, $f_\mathrm{d(obs)}$, $f_\mathrm{d(sim)}$ and $d$ as described in Section \ref{mwti}. The combination of feedback parameters that produces the lowest value of $d$, i.e., the most realistic galaxy in terms of its baryonic makeup is highlighted for each setup. We have included the sections in which each individual simulation is discussed, in order to guide the reader.}
	\label{tab:fbpara}
	\begin{tabular}{|p{.3\columnwidth}|p{.19\columnwidth}|p{.17\columnwidth}|p{.17\columnwidth}|p{.17\columnwidth}|p{.19\columnwidth}|p{.18\columnwidth}|p{.29\columnwidth}|}
		\hline
		\multicolumn{8}{|c|}{Feedback setup list} \\
		\hline
		\multicolumn{8}{|c|}{MW galaxy zoom simulations -- Setup 1} \\
		\hline
		Feedback parameters & $m_{500}\  [M_\odot]$ & $f_\mathrm{d(obs)}$ & $f_\mathrm{d(sim)}$ & $f_\mathrm{s(obs)}$ & $f_\mathrm{s(sim)}$ & $d$ & Discussed in Section\\
		\hline
		(1.0\e{-5}, 1\_3, 0.1) & 5.59\e{11} & 0.26 & 1.41 & 0.22 & 1.25 & 1.54 & \ref{mwt}\\
		(1.0\e{-4}, 1\_3, 0.1) & 4.23\e{11} & 0.24 & 0.68 & 0.20 & 0.57 & 0.57 & \ref{mwt}\\
		(2.5\e{-4}, 1\_3, 0.1) & 4.92\e{11} & 0.25 & 0.82 & 0.21 & 0.42 & 0.60 & \ref{mwt}\\
		(5.0\e{-4}, 1\_3, 0.1) & 3.99\e{11} & 0.24 & 0.68 & 0.20 & 0.57 & 0.57 & \ref{mwt}\\
		(1.0\e{-3}, 1\_3, 0.1) & 3.88\e{11} & 0.24 & 0.19 & 0.19 & 6.01\e{-3} & 0.19 & \ref{mwt}\\
		(1.0\e{-5}, 2\_6, 0.1) & 4.65\e{11} & 0.25 & 1.02 & 0.20 & 0.91 & 1.04 & \ref{mwt}\\
		(1.0\e{-5}, 3\_9, 0.1) & 4.30\e{11} & 0.24 & 0.81 & 0.20 & 0.70 & 0.76 & \ref{mwt}\\
		(1.0\e{-5}, 4\_12, 0.1) & 5.05\e{11} & 0.25 & 0.86 & 0.21 & 0.54 & 0.69 & \ref{mwt}\\
		(1.0\e{-5}, 5\_15, 0.1) & 5.66\e{11} & 0.26 & 1.26 & 0.22 & 0.15 & 1.00 & \ref{mwt}\\
		(1.0\e{-5}, 1\_3, 0.2) & 5.53\e{11} & 0.26 & 1.37 & 0.22 & 1.22 & 1.49 & \ref{mwt}\\
		\rowcolor{lightgray} (2.5\e{-4}, 1\_3, 0.2) & 4.05\e{11} & 0.24 & 0.36 & 0.19 & 0.18 & 0.13 & \ref{mwt}\\
		\hline
		\multicolumn{8}{|c|}{MW galaxy zoom simulations -- Setup 2} \\
		\hline
		Feedback parameters & $m_{500}\  [M_\odot]$ & $f_\mathrm{d(obs)}$ & $f_\mathrm{d(sim)}$ & $f_\mathrm{s(obs)}$ & $f_\mathrm{s(sim)}$ & $d$ & Discussed in Section\\
		\hline
		(2.5\e{-4}, 1\_3, 0.2) & 3.67\e{11} & 0.23 & 0.23 & 0.18 & 5.64\e{-3} & 0.17 & \ref{mwti}\\
		(2.5\e{-4}, 1\_3, 0.9) & 3.65\e{11} & 0.23 & 0.12 & 0.18 & 2.76\e{-3} & 0.21 & \ref{mwti}\\
		(5.0\e{-5}, 1\_3, 0.9) & 3.46\e{11} & 0.23 & 0.33 & 0.18 & 4.87\e{-2} & 0.16 & \ref{mwti}\\
		(4.0\e{-5}, 1\_3, 0.9) & 3.47\e{11} & 0.23 & 0.28 & 0.18 & 9.08\e{-2} & 0.11 & \ref{mwti}\\
		(3.0\e{-5}, 1\_3, 0.9) & 3.69\e{11} & 0.23 & 0.44 & 0.18 & 0.13 & 0.22 & \ref{mwti}\\
		(2.0\e{-5}, 1\_3, 0.9) & 4.15\e{11} & 0.24 & 0.61 & 0.19 & 0.39 & 0.42 & \\
		\rowcolor{lightgray} (3.0\e{-5}, 1\_1, 0.9) & 3.50\e{11} & 0.23 & 0.24 & 0.18 & 0.10 & 7.59\e{-2} & \ref{mwti} and \ref{chaos}\\
		\rowcolor{lightgray} (2.5\e{-5}, 1\_1, 0.9) & 3.71\e{11} & 0.23 & 0.29 & 0.18 & 0.14 & 7.3\e{-2} & \ref{mwti} and \ref{chaos}\\
		(3.0\e{-5}, 1\_1, 0.9) run 2 & 3.78\e{11} & 0.23 & 0.32 & 0.18 & 8.34\e{-2} & 0.14 & \ref{chaos}\\
		(2.5\e{-5}, 1\_1, 0.9) run 2 & 3.59\e{11} & 0.23 & 0.33 & 0.18 & 0.15 & 0.10 & \ref{chaos}\\
		\hline
		\multicolumn{8}{|c|}{Dwarf galaxy zoom simulations - Setup 2} \\
		\hline
		Feedback parameters & $m_{500}\  [M_\odot]$ & $f_\mathrm{d(obs)}$ & $f_\mathrm{d(sim)}$ & $f_\mathrm{s(obs)}$ & $f_\mathrm{s(sim)}$ & $d$ & Discussed in Section\\
		\hline
		(3.0\e{-5}, 1\_1, 0.9) & 2.34\e{10} & 9.38\e{-2} & 8.14\e{-2} & 2.60\e{-2} & 8.47\e{-3} & 2.15\e{-2} & \ref{dwarfti}\\
		(2.5\e{-5}, 1\_1, 0.9) & 2.33\e{10} & 9.36\e{-2} & 0.12 & 2.58\e{-2} & 9.46\e{-3} & 2.79\e{-2} & \ref{dwarfti}\\
		\hline
	\end{tabular}
\end{table*}

\subsection{MW galaxy zoom simulations with Setup 2}\label{mwti}

We perform the following parameter space exploration with Setup 2 in Table \ref{tab:sfsetup}. With this setup, we run a total of 49 simulations in order to calibrate the feedback prescription, and we make a similar classification as before, shown in Figure \ref{fig:t_para}. We summarise the various properties of the halo of interest of simulations with Setup 1 and 2 in Table \ref{tab:fbpara}. This table includes simulations that will be discussed in Sections \ref{dwarfti} and \ref{chaos}.

{\parskip 2em} From the 49 simulations, only one simulation with (3.0\e{-5}, 1\_1, 1.0) failed to reach $z=0$ due to the complete removal of gas when stars form. The process of iteration started from the best combination of parameters found in Section \ref{mcgaugh_time_sim}, (2.5\e{-4}, 1\_3, 0.2) and progressed based on the trends found in Figure \ref{fig:trend} to move the simulation data point closer to the target. This process will be explained later. We introduce a measure of closeness between the simulated and the observed galaxy properties via the Cartesian distance to the target,
\begin{equation}
d = \sqrt{(f_{\mathrm{s(sim)}} - f_{\mathrm{s(obs)}})^2 + (f_{\mathrm{d(sim)}} - f_{\mathrm{d(obs)}})^2}, 
\label{eq:fom}
\end{equation} where subscripts $sim$ and $obs$ refers to simulation and observation respectively. Lower values of $d$ represent a more realistic simulated galaxy in terms of both $f_s$ and $f_d$. For the goodness of fit of individual properties, we refer to Table \ref{tab:fbpara}. 

{\parskip 2em} Comparing the feedback parameter values covered in both Setup 1 and 2, it is clear that they do not cover an equal area of parameter space. The main differences lie in the usage of high $f_*$ while having low values of $r$ and $\epsilon$ in Setup 2 as compared to Setup 1. There are two significant volumes of parameter space not covered in Setup 2: large values of $\epsilon$ coupled with low $r$ and $f_*$ and large values of $r$ with low values of $\epsilon$ and $f_*$. Also, there are regions (intermediate values of $\epsilon$ and $f_*$, high values of $r$ and intermediate values of $f_*$) in the parameter space of Setup 2 that are not sampled. The reason why we do not have any simulations in these regions will be explained in the next section with Figure \ref{fig:ti_trend}.

\subsubsection{Comparison to Baryonic properties from \citet{2010ApJ...708L..14M} -- Setup 2}\label{mcgaugh_ti_sim}

We will identify the best star formation and feedback parameters through an iterative process beginning from the initial point (2.5\e{-4}, 1\_3, 0.2) from before, applying the knowledge of trends from Figure \ref{fig:trend}. We use arrows to represent the general movement of data points due to the initial adjustments of $f_*$ and $\epsilon$ before using $r$ and $s$ for the finer last adjustments on the $f_\mathrm{s}$ and $f_\mathrm{d}$ plane. We present this with a representative set of simulations in Figure \ref{fig:ti_trend}, similar to Figure \ref{fig:trend} by starting from the best combination of parameters (blue dot) in Setup 1. It is evident that identical feedback prescription in different settings produced a MW with disparate $f_s$ and $f_d$. In Setup 2, the previously optimal values produced a MW galaxy with minimal stellar mass. This small amount of stars at $z=0$ is a result of the relaxed star formation conditions producing numerous small star formation events, which instantly yield feedback and reduces future star formation.

{\parskip 2em} From the starting point, we increase $f_*$ from 0.2 to 0.9 (see green arrow in Figure \ref{fig:ti_trend}). This trend indicates that as $f_*$ increases, $f_\mathrm{d}$ decreases while $f_\mathrm{s}$ stays constant, which is in agreement with the combination of effects of the green and blue arrows shown in Figure \ref{fig:trend}. Despite only having two data points, we know from the direction given by the green arrow in Figure \ref{fig:trend} that it will have the same effect on the properties as increasing $\epsilon$ (blue arrow). Therefore, if we increase $f_*$ further in Figure \ref{fig:trend}, we can expect it to follow the last blue arrow, which is a horizontal motion of decreasing $f_\mathrm{d}$ with constant $f_\mathrm{s}$. Together with the immediate feedback from stars, increasing $f_*$ converts more gas into stars, which reduces the amount of gas, leading to the decline in $f_\mathrm{d}$. Although more stars form initially, the feedback is stronger, reducing the amount of gas available to form more stars as the halo aged, resulting in a constant $f_\mathrm{s}$. Therefore, we increase $f_*$ in an attempt to move the data point as far left as possible in Figure \ref{fig:ti_trend} in preparation for the next step. The simulation with $f_*=1.0$ does not produce a MW galaxy with significantly different $f_\mathrm{s}$ and $f_\mathrm{d}$. Furthermore, this value of $f_*$ caused the only failed run from 49 simulations. Hence, we settle on a $f_*$ value of 0.9 (orange dot) as the starting point for the next phase of iteration. 

{\parskip 2em} After obtaining the minimal $f_\mathrm{d}$ with (2.5\e{-4}, 1\_3, 0.9), we attempt to increase $f_\mathrm{s}$ and $f_\mathrm{d}$ in the next iteration to move closer to the target. From what we have learned from Figure \ref{fig:trend}, we can achieve this by either decreasing $\epsilon$ or $r$. Since $r$ is already at a minimum, lowering $\epsilon$ is the only option. We present only a representative set of data points connected by the blue arrows to illustrate the general change in $f_\mathrm{s}$ and $f_\mathrm{d}$ due to smaller $\epsilon$ values. This increase in $f_\mathrm{s}$ and $f_\mathrm{d}$ is in agreement with Figure \ref{fig:trend}, explained by the less efficient baryon expulsion, which leads to higher star formation and retention of gas within $r_{500}$.

{\parskip 2em} The final step is to adjust $r$ and $s$ to improve the match to the observed $f_\mathrm{s}$ and $f_\mathrm{d}$. Initially, we maintain the injection of feedback energy in a cube and increase the size, i.e, from $r$ = 1 and $s$ = 3 to $r$ = 2 and $s$ = 6. The aim is to obtain a point to the top right of the target and increase $r$ and $s$ correspondingly to move it towards the target as predicted by Figure \ref{fig:trend}. However, we do not obtain any good fit. Coupled with an upper limit to the extent of feedback injection where $f_\mathrm{d}$ increases instead beyond $r$ = 3 and $s$ = 9 (see Figure \ref{fig:trend}), we decide to change the shape of energy injection from a cube to just the adjacent cells centred around the star particle. In parameters terms, we change $r$ = 1 and $s$ = 3 to $r$ = 1 and $s$ = 1. As a result, the feedback energy is injected into four instead of 27 cells, effectively increasing the energy concentration per cell by approximately an order of magnitude. This increased energy density causes a larger decrease in $f_\mathrm{d}$ than in $f_\mathrm{s}$. In contrast, increasing the extent of feedback injection maintained in a cube region generates a comparable change in both $f_\mathrm{s}$ and $f_\mathrm{d}$.  

\begin{figure*}
	\includegraphics[width=0.80\linewidth]{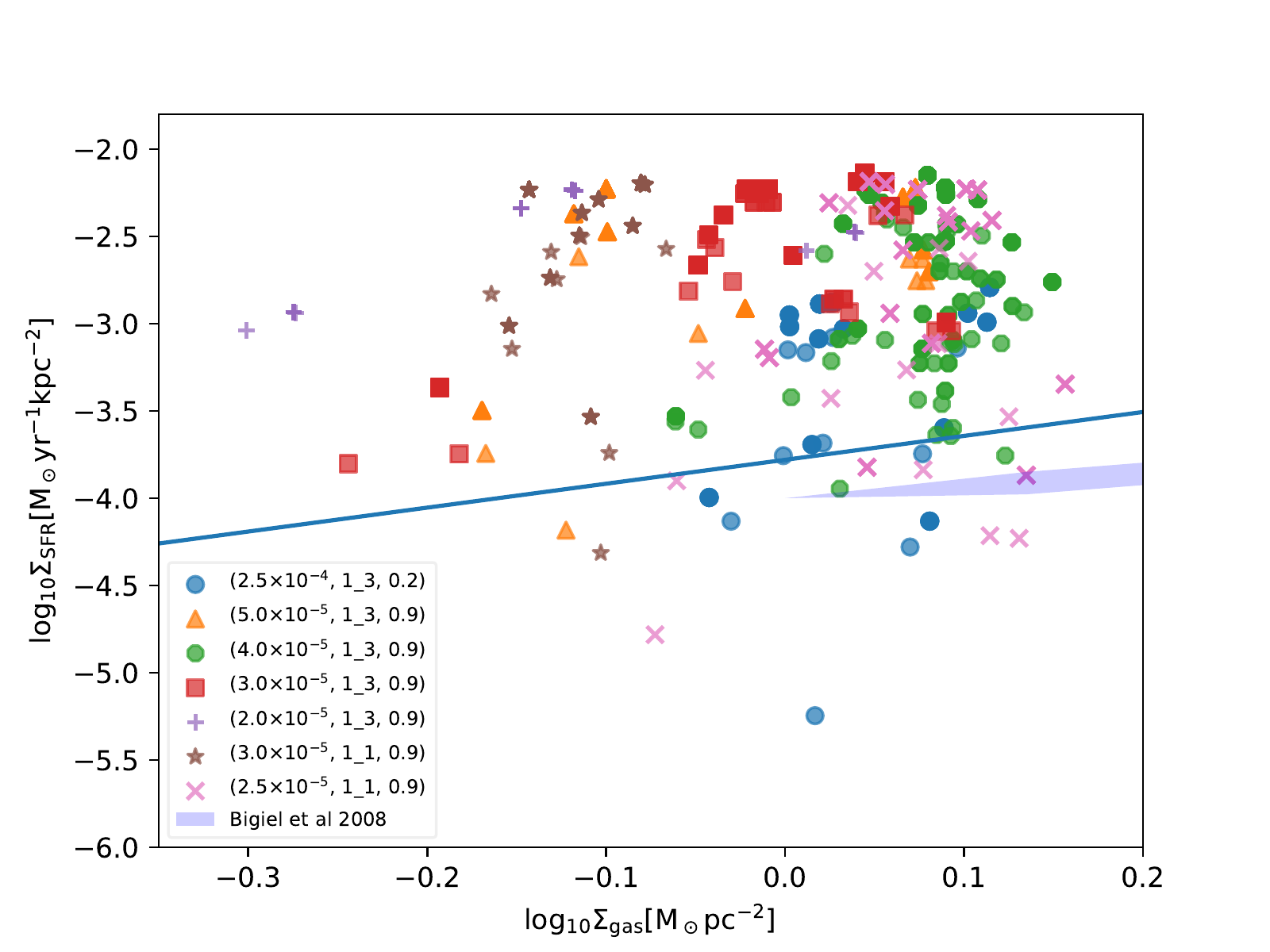}
	\caption{Graph of SFR surface density against gas surface density illustrating the KS relation as in Figure \ref{fig:ks_t}. Different coloured points are simulation data with sub-kpc resolution consistent with rough approximation from the observations in nearby galaxies by \citet{2008AJ....136.2846B} represented by the blue hatched contours. The blue line is derived from the observational fit of \citet{2007ApJ...671..333K}. As a result of the difference in the feedback prescription, the simulated galaxy have a much lower gas surface density as compared to Figure \ref{fig:ks_t}. For further description of this figure, we refer to Section \ref{ksti}.}
	\label{fig:ks_ti}
\end{figure*}

{\parskip 2em} We determine (2.5\e{-5}, 1\_1, 0.9) and (3.0\e{-5}, 1\_1, 0.9) as the two sets of parameters able to produce the smallest $d$ value (see Table \ref{tab:fbpara}). Given the vast area of unexplored parameter space and the starting point of the iterative process, we justify that the steps taken constitute the most reasonable route through parameter space that can produce a close match to observations. The starting values of (2.5\e{-4}, 1\_3, 0.2) define the boundaries where values can be adjusted. $r$ and $f_*$ are almost at the minimum, meaning they can only increase while $\epsilon$ can either decrease or increase. Furthermore, the low $f_{\rm s}$ of the starting point of properties in Figure \ref{fig:ti_trend} suggests that the current feedback is too strong that it restricts star formation. 

{\parskip 2em} Together with the trends of changing parameters, the possible motions of the data point are a horizontal movement to the left or right, and diagonally right. The worst possible option is to increase $\epsilon$, moving the data point to the left. This choice leaves us stranded because we cannot create further motion since $r$ and $f_*$ are already close to their minimum values. The next possible option is to increase $r$ above 3, causing the data to move horizontally right. The next steps associated with this first movement will be decreasing $\epsilon$ to iterate data points towards the top right before increasing $f_*$ to reduce the data to match the target. However, given the initial movement away from the target, we believe that this will not produce a better match than what is presented. The most plausible option is to decrease $\epsilon$, moving the data point along the blue arrows indicated in Figure \ref{fig:ti_trend}. $f_*$ can then be increased to move it down diagonally left towards the target while fine-tuning $r$ and $s$. This change is preferred over increasing $r$ because of the turn around expected beyond $r=4$, which limits the degrees of freedom. However, following this option will generate a combination of parameters similar to what we have found. Out of the possible options to move the initial point in parameter space, we have chosen the path that will produce the best match to fit the observational data from \citet{2010ApJ...708L..14M}. Since the argument put forth does not mention the possibility of an ideal set of parameters lying in the region of parameter space consisting of intermediate values of $\epsilon$ and $f_*$ and high values, they are not investigated.

{\parskip 2em} Comparing the values of the feedback parameters that reproduce the MW baryonic makeup from both setups, we can identify the self-consistency of our feedback implementation. Setup 1 yielded an optimal combination of (2.5\e{-4}, 1\_3, 0.2) but in Setup 2, we conclude that (2.5\e{-5}, 1\_1, 0.9) and (3.0\e{-5}, 1\_1, 0.9) reproduced the most realistic MW galaxy. In Setup 2, the simulation forms more star particles but they are of lower masses than in Setup 1. Therefore, in order to produce a similar amount of stars observed in a MW galaxy at $z=0$, Setup 2 requires a higher gas to star conversion efficiency, 0.9 as compared to 0.2 in Setup 1. In response to this larger conversion efficiency, Setup 2 require a lower $\epsilon$. The value of $\epsilon$ differs significantly between the setups as a result. Setup 2 is preferred because of the more realistic star formation history in the dwarf galaxy (see Section \ref{dwarft}), and a more extensive exploration of parameter space due to higher computational resources efficiency.

\subsubsection{Kennicutt--Schmidt relation -- Setup 2} \label{ksti}

In this section, we will present the agreement of star formation in the simulation with the KS relation described in Section \ref{ks-relation}. As in Figure \ref{fig:ks_t}, we choose non-zero SFR patches within $r_{500}$ at $z=0$ and compare it to the fit given by Equation \ref{eq:ks_fit} and observations of nearby galaxies by \citet{2008AJ....136.2846B}, shown in Figure \ref{fig:ks_ti}. There is a clustering of points around the fit but no slope can be deduced from the points. Also, the simulated gas density is too low for comparison to observational data. We believe the concentration of points around low surface gas density is due to the relaxed star formation criteria and the higher $f_*$. These conditions result in a more efficient conversion of gas into stars, leading to more feedback energy injection that lowers the gas density. 

{\parskip 2em} While (3.0\e{-5}, 1\_1, 0.9) and (2.5\e{-5}, 1\_1, 0.9) recovers $f_\mathrm{s}$ and $f_\mathrm{d}$ well, there is an absence of patches with high gas surface density, restricting our ability to probe the KS relation in that regime. This absence also suggests that feedback might have been too efficient in driving gas out of the central region of the galaxy. Comparing Setup 2 to Setup 1, the former is not as good in recovering the KS relation. Setup 2 provides a relatively more instantaneous conversion of gas into stars, which drives gas surface density to lower values. As discussed earlier, a larger quantity of stars is formed in Setup 2, which begins feeding back into the IGM immediately. Coupled with the high conversion efficiency of gas to stars, it empties the central region of the galaxy of gas, explaining why the gas surface density is low. 

\subsubsection{Haloes in the high-resolution region -- Setup 2} \label{sat_ti}

\begin{figure*}	
	\begin{subfigure}[b]{1.0\columnwidth}
		\includegraphics[width=\linewidth]{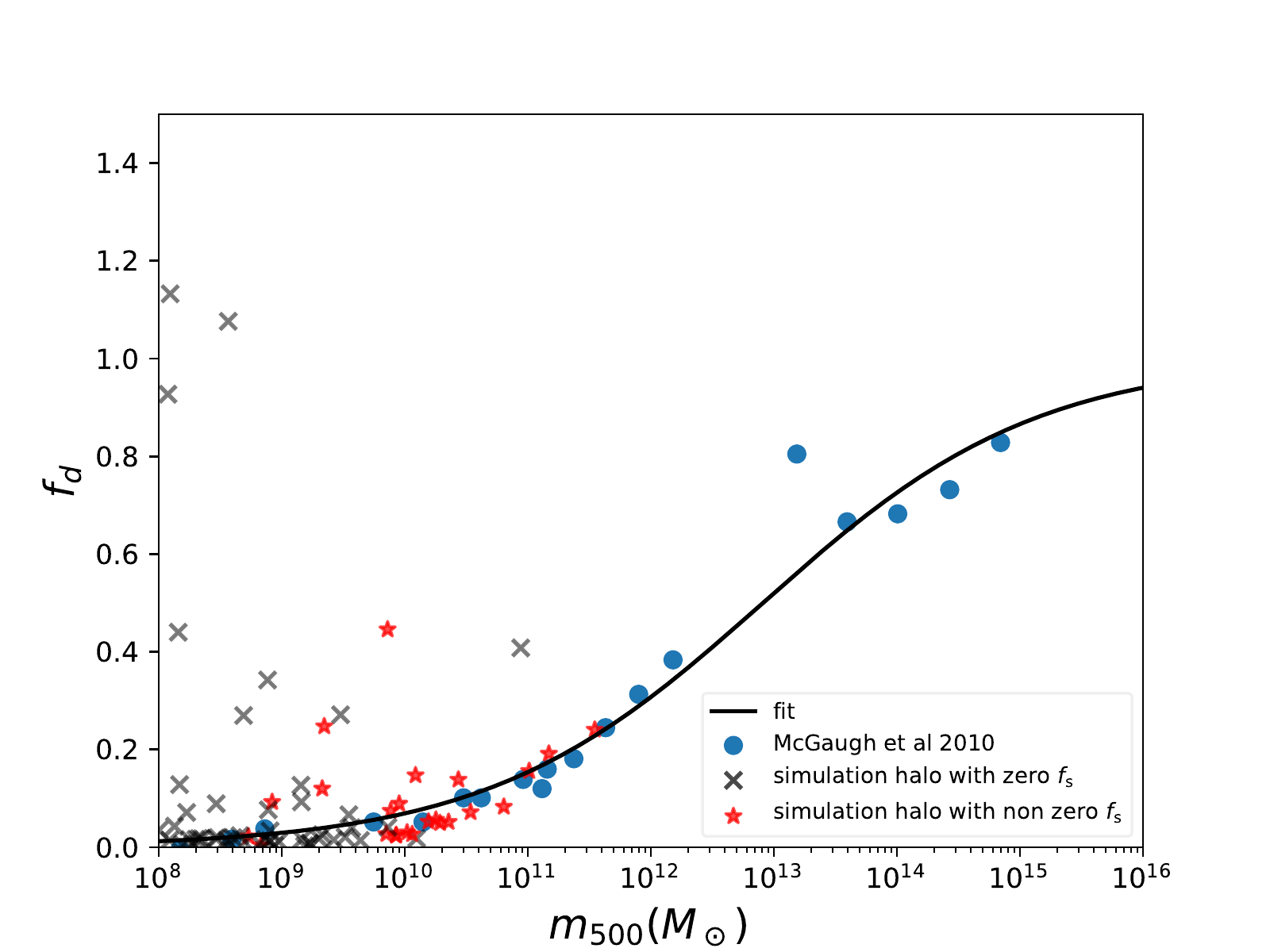}
		\label{fig:ti_sat1_fd}
	\end{subfigure}
	\hfill 
	\begin{subfigure}[b]{1.0\columnwidth}
		\includegraphics[width=\linewidth]{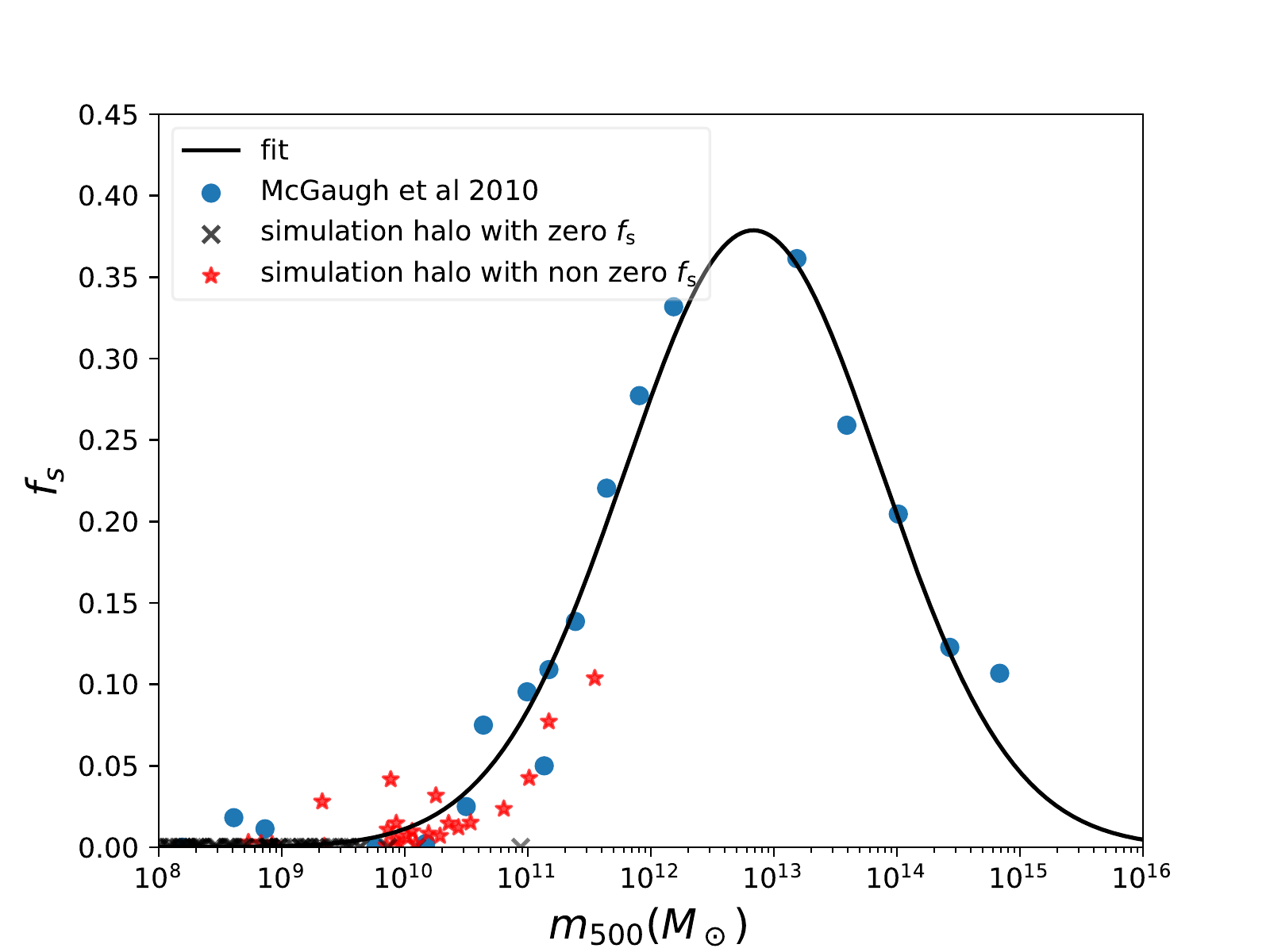}
		\label{fig:ti_sat1_fs}
	\end{subfigure}
	\begin{subfigure}[b]{1.0\columnwidth}
		\includegraphics[width=\linewidth]{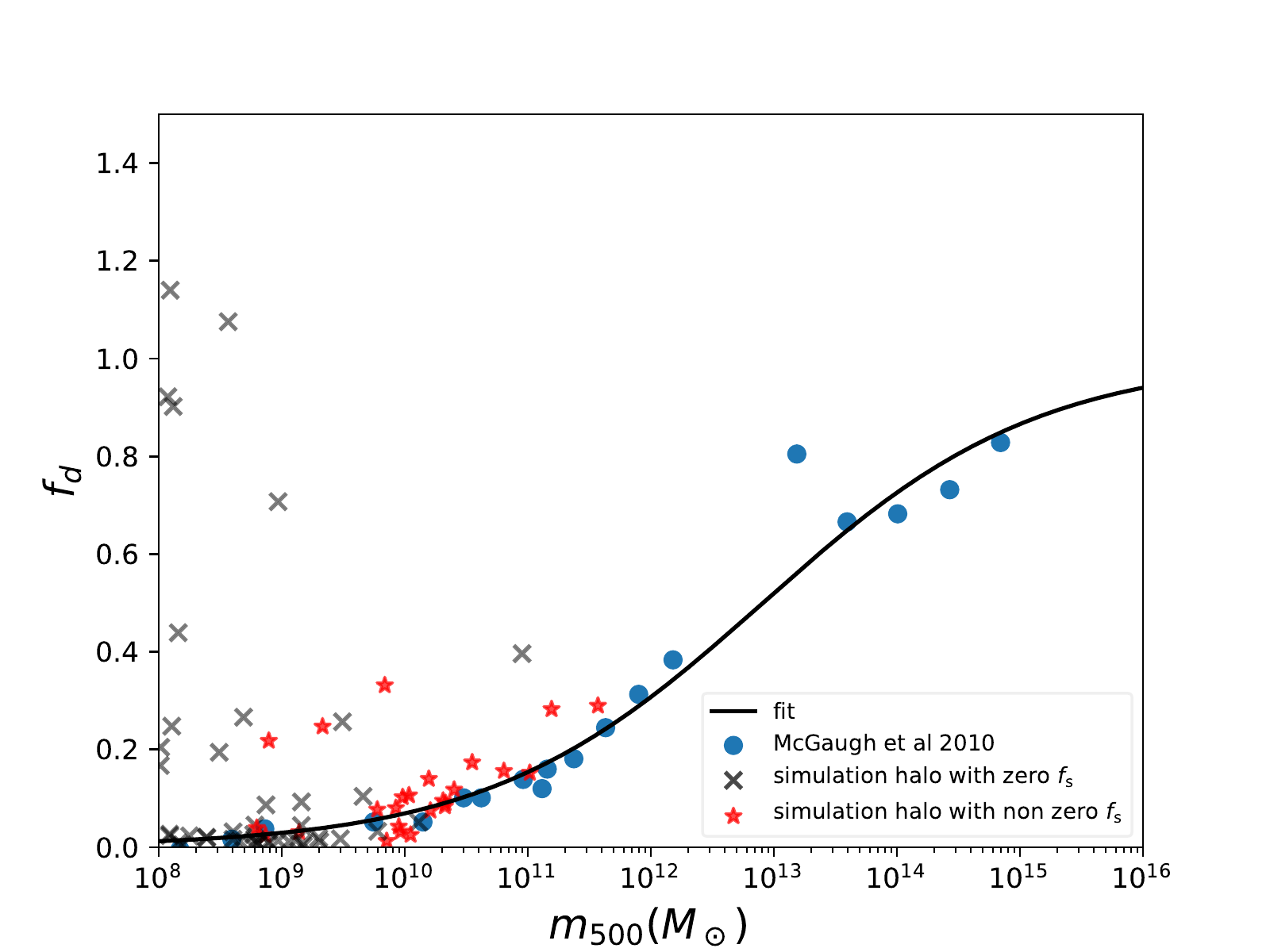}
		\label{fig:ti_sat2_fd}
	\end{subfigure}
	\hfill 
	\begin{subfigure}[b]{1.0\columnwidth}
		\includegraphics[width=\linewidth]{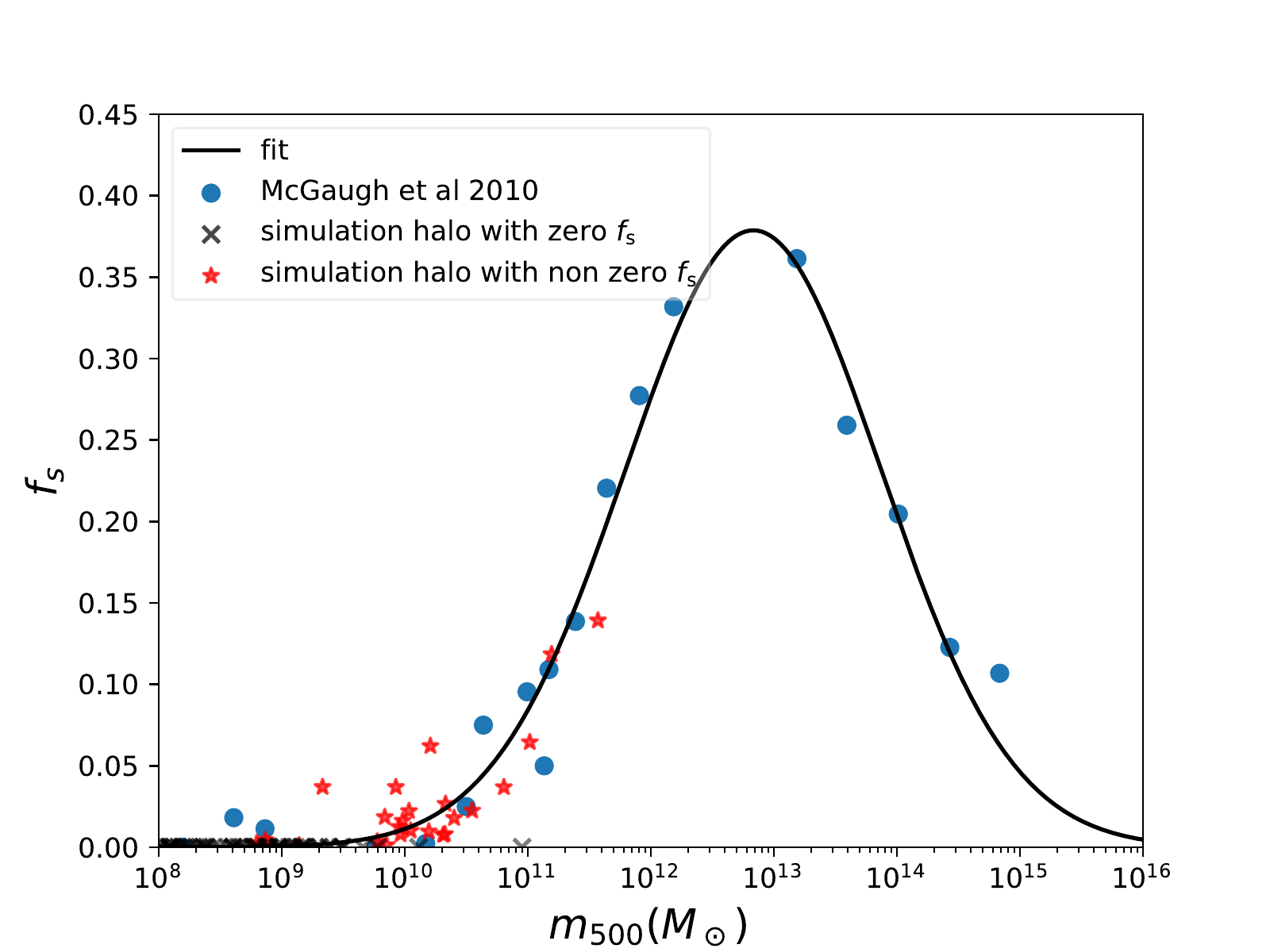}
		\label{fig:ti_sat2_fs}
	\end{subfigure}
	\caption{Graph of $f_\mathrm{d}$ against $m_{500}$ (left) and $f_\mathrm{s}$ against $m_{500}$ (right) with equations described in Section \ref{mcgaugh}. The black line represents the fit given by Equation \ref{eq:fd_fit} and \ref{eq:fs_fit}. The blue dots are data from \citet{2010ApJ...708L..14M} while the crosses and stars are properties of haloes with various mass from the high-resolution region in the simulation. The black cross and red star refer to haloes in the must refine region with zero and non zero $f_\mathrm{s}$ respectively. Top row figures are from (3.0\e{-5}, 1\_1, 0.9) while the bottom row figures are from (2.5\e{-5}, 1\_1, 0.9). In contrast to the limited success illustrated in Setup 1 (see Figure \ref{fig:mcgaugh_t_sat}), these feedback prescriptions in Setup 2 are able to reproduce the baryonic makeup in haloes between $10^{10}\,\mathrm{M_\odot}$ and $10^{12}\,\mathrm{M_\odot}$. Refer to Section \ref{sat_ti} for a detailed description.}
	\label{fig:mcgaugh_ti_sat}
\end{figure*}

As in Section \ref{sat_t}, we look at the $f_\mathrm{s}$ and $f_\mathrm{d}$ of the other haloes within the high-resolution region of three virial radii from the MW halo. We plot $f_\mathrm{d}$ against $m_{500}$ on the left column, $f_\mathrm{s}$ against $m_{500}$ on the right column, and simulations with (3.0\e{-5}, 1\_1, 0.9) and (2.5\e{-5}, 1\_1, 0.9) on the top and bottom rows in Figure \ref{fig:mcgaugh_ti_sat} respectively.

{\parskip 2em} With the exception of one and two haloes from the runs with (3.0\e{-5}, 1\_1, 0.9) and (2.5\e{-5}, 1\_1, 0.9) respectively, we find very good agreement for both $f_\mathrm{s}$ and $f_\mathrm{d}$ of haloes between $10^{10}\,\mathrm{M_\odot}$ and $10^{12}\,\mathrm{M_\odot}$. This agreement is in contrast to Figure \ref{fig:mcgaugh_t_sat} where agreement is only achieved for $f_\mathrm{s}$ and not $f_\mathrm{d}$. On top of that, the level of agreement with observations is much better in Figure \ref{fig:mcgaugh_ti_sat} than Figure \ref{fig:mcgaugh_t_sat} as points lie closer to the fit. For haloes below $10^{10}\,\mathrm{M_\odot}$, it is plausible that the lack of mass and spatial resolution is the cause of their inability to form stars. On the other hand, the larger mass haloes that suffer the same problem require future zoom simulations to be carried out in order to identify the root of the issue. 

\subsection{Dwarf galaxy zoom simulation with Setup 2} \label{dwarfti}

\begin{figure}
	\includegraphics[width=\linewidth]{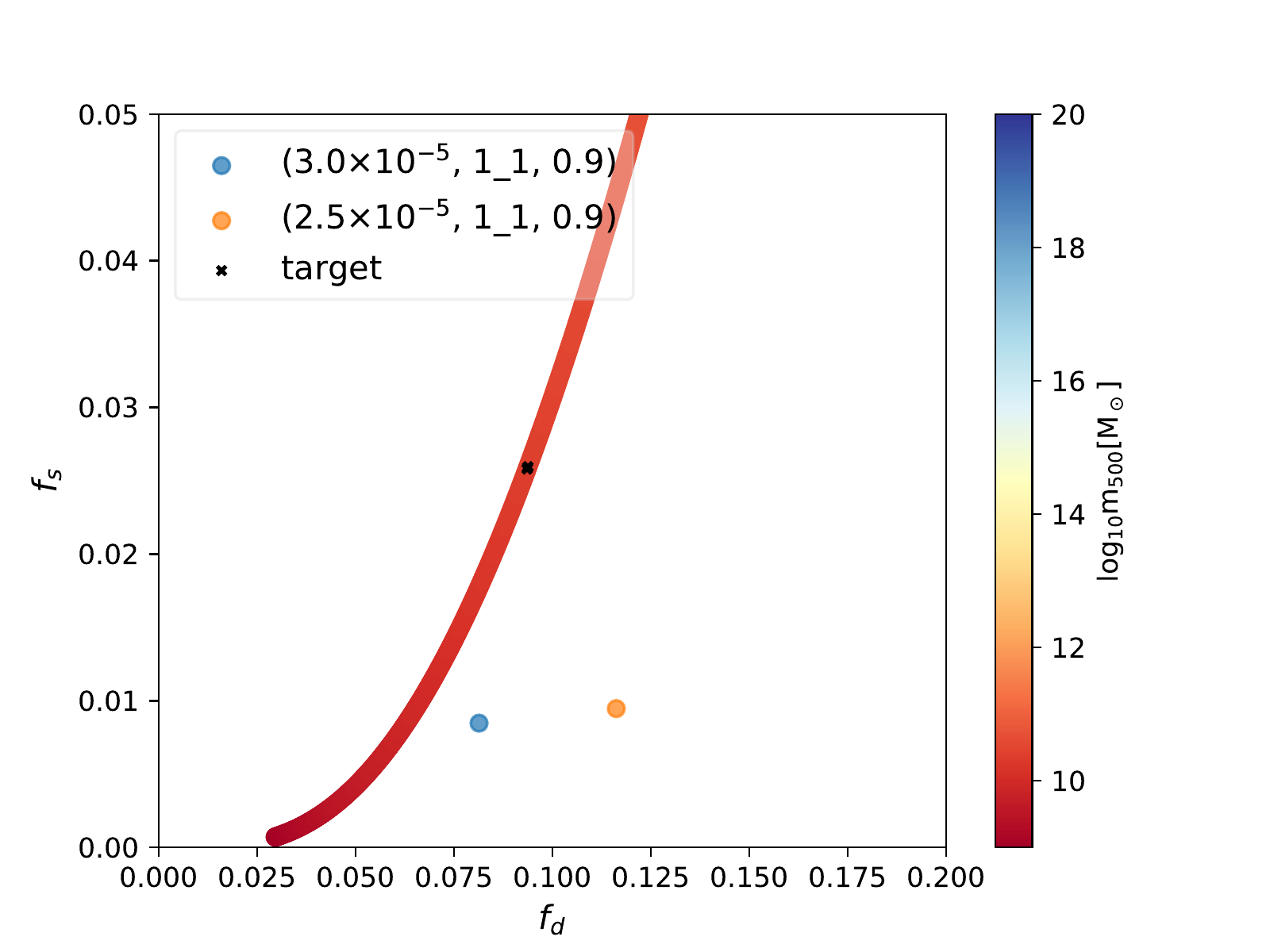}
	\caption{Plot of $f_\mathrm{s}$ against $f_\mathrm{d}$ for the zoom dwarf galaxy simulations. Various coloured dots represent runs with different set of feedback parameters with the colour bar having the usual meaning. It is focused on a small area near the target due to closeness of simulation results with the observed properties. Consistent with Figure \ref{fig:mcgaugh_ti_sat}, (3.0\e{-5}, 1\_1, 0.9) and (2.5\e{-5}, 1\_1, 0.9) are able to produce a dwarf galaxy with $f_\mathrm{s}$ and $f_\mathrm{d}$ close to observations. See Section \ref{dwarfti} for a detailed description.}
	\label{fig:mcgaugh_sh}
\end{figure}

We conduct zoom simulations of a dwarf galaxy with $m_\mathrm{vir}$ of approximately $10^{10}\,\mathrm{M_\odot}$ as an additional test of the universality of the feedback parameters in different halo mass bins. We described how we pick this dwarf galaxy from the high-resolution region of the MW zoom simulation in Section \ref{simsetup}. Similarly, we increase the number of nested levels to keep the number of particles defining the halo constant with that of the MW while keeping the spatial resolution constant. We then compare the $f_\mathrm{s}$ and $f_\mathrm{d}$ of the halo to \citet{2010ApJ...708L..14M} in Figure \ref{fig:mcgaugh_sh}.

{\parskip 2em} We present a close-up view of the parameter space in Figure \ref{fig:mcgaugh_sh} because we are showing results from zoom simulations of the dwarf galaxy using the two best sets of parameters only. It is clear that the $f_\mathrm{s}$ and $f_\mathrm{d}$ of the simulated galaxy in both feedback prescriptions are comparable to the target. We expect good agreement based on the results of Figure \ref{fig:mcgaugh_ti_sat}. Therefore, we argue that this feedback prescription is insensitive to mass resolution with a smaller mass halo having a lower and higher resolution in Figures \ref{fig:mcgaugh_ti_sat} and \ref{fig:mcgaugh_sh} respectively. However, it is also essential to investigate the dependence of the feedback prescription on spatial resolution in future work. 

\subsection{Chaos and variance} \label{chaos}

\begin{figure}
	\includegraphics[width=\linewidth]{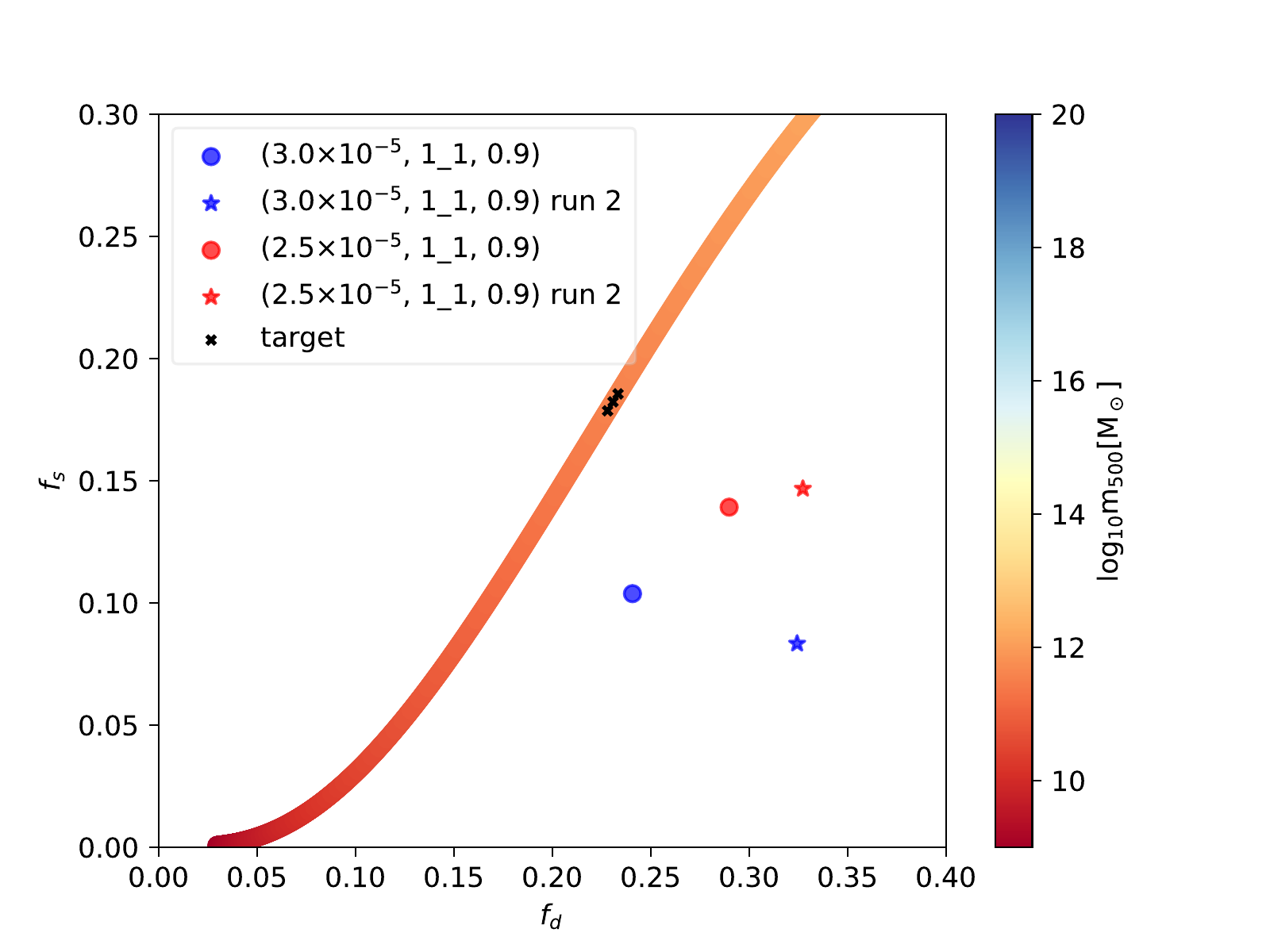}
	\caption{Plot of $f_\mathrm{s}$ against $f_\mathrm{d}$ for pairs on zoom MW simulation with identical initial conditions and feedback prescriptions evolved with different processors. The same coloured symbols refer to the simulations with identical setup while dots and stars represent identical simulations with two different sets of processors. The colour bar has its usual meaning. It is again focused in a small area near the target due to the level of agreement of simulation results with the properties. Rerunning a simulation with identical initial conditions can produce simulated properties that differ significantly. See Section \ref{chaos} for a detailed description.}
	\label{fig:chaos}
\end{figure}

Recognising the argument put forth by \citet{2019MNRAS.482.2244K} for chaotic variance in numerical simulations, we conduct our zoom simulations twice on different processors. They have identical initial conditions and feedback prescriptions but evolved on different combinations of processors in the same computing cluster. The aim is to find out how much the halo properties would differ from each other due to the usage of a different set of processors. We quantify this difference in Figure \ref{fig:chaos}.

{\parskip 2em} Dots and stars in Figure \ref{fig:chaos} represent the pair of simulations with (3.0\e{-5}, 1\_1, 0.9) (blue) and (2.5\e{-5}, 1\_1, 0.9) (red) respectively. Despite both of them being close to the target, $f_\mathrm{s}$ and $f_\mathrm{d}$ for each pair can differ as much as running a simulation with a different set of feedback parameters. Comparing (3.0\e{-5}, 1\_1, 0.9) run 2 to (4.0\e{-5}, 1\_1, 0.9) in Figure \ref{fig:ti_trend}, the simulated galaxies have similar values of $f_\mathrm{s}$ and $f_\mathrm{d}$. This variance is also apparent from the values of $m_\mathrm{500}$ where the maximum, minimum and the mean values are shown by the black crosses.

{\parskip 2em} Looking at Figure \ref{fig:chaos}, the deviation in $f_\mathrm{s}$ from the pair of simulations is comparable to the $10\%$ difference in stellar mass concluded in \citet{2019MNRAS.482.2244K} despite not using identical processors. However, the deviation in total baryon mass is as high as $33\%$, possibly arising from the coupling of star formation and feedback where a 10\% difference in stellar mass affects the feedback significantly. There is not a consistent trend observed in Figure \ref{fig:chaos}, i.e., increase or decrease in $f_s$ both cause an increase in $f_d$. We attribute this to these ratios containing a mixture of stellar and gas mass. Due to the complex coupling of star formation and feedback, it is difficult to disentangle the contribution of each component. For example, increasing stellar mass results in a decrease in gas mass but it is unclear which is the more dominant effect. As a result, the baryonic composition of the halo can differ drastically. 
 
\section{Summary and Discussion}\label{discussion}
We present results from a large number of zoom simulations of both a MW and a dwarf galaxy. This suite of simulations is the first application of numerical simulations calibrated to match the baryon content and stellar fraction properties presented by \citet{2010ApJ...708L..14M}. Using the star formation routine of \citet{1992ApJ...399L.113C} and the thermal supernova feedback of \citet{2006ApJ...650..560C}, we select factors such as $f_*$ to tune the conversion efficiency of gas to stars, $\epsilon$ for the feedback energy budget, and lastly, both $r$ and $s$ to calibrate the extent of feedback injection in the simulations. We also identify additional parameters that require adjustments in order to achieve realistic star formation histories. They are the Jeans instability check, the star particle threshold mass and the timestep dependence of star formation. These directly influence the criteria used to determine the occurrence of star formation.

{\parskip 2em} It is remarkable that there is such a small variance associated with the data presented by \citet{2010ApJ...708L..14M}. This is the main reason why we strive to improve the agreement between our simulation results and observations as much as possible. However, it is also important to note the possibility of underestimates in the errors and unaccounted systematics. The method of determining the mass of the halo from observations affects the amount of scatter too. If abundance matching is used, $m_*$ will have a lot more scatter than $m_b$ in the Tully-Fisher plane at low mass, leading a corresponding amount of scatter in $f_s*$ and $f_d$. Since most of the mass in low mass rotating galaxies is gas and not stars, one can also question the applicability of extrapolating abundance matching relations to such low masses.

{\parskip 2em} With the mentioned parameters, we produce a MW galaxy with realistic baryon and stellar fraction when compared to the observations of \citet{2010ApJ...708L..14M} with our suite of simulations. We achieve this agreement with two different setups shown in Table \ref{tab:sfsetup}. Setup 1 utilises a timestep dependent star formation with Jeans instability check and a star formation threshold mass of $10^{5}\,\mathrm{M_\odot}$. We attempt a total of 22 simulations with this setup and find that (2.5\e{-4}, 1\_3, 0.2) managed to reproduce the observed $f_\mathrm{s}$ and $f_\mathrm{d}$. However, the simulated MW galaxy in this feedback prescription does not match the observed KS relation very well. By applying this feedback prescription to a zoom simulation of a dwarf in this setup, we find star formation starting too late as compared to the simulated MW galaxy. To resolve this issue, we propose switching to a timestep independent star formation setup with no Jeans instability check and threshold mass (Setup 2). However, due to the non-linear coupling of the various processes in the simulation, a new prescription requires re-exploration of subgrid parameters.

{\parskip 2em} We begin an iterative process from (2.5\e{-4}, 1\_3, 0.2) in Setup 2, concluding with two sets of parameters that produced a close fit to the $f_\mathrm{s}$ and $f_\mathrm{d}$ with the use of 49 simulations. They are (2.5\e{-5}, 1\_1, 0.9) and (3.0\e{-5}, 1\_1, 0.9). As in Setup 1, there are issues with the KS relation of the simulated galaxy. However, these feedback prescriptions performed remarkably well in matching the baryonic makeup of haloes between $10^{10}\,\mathrm{M_\odot}$ and $10^{12}\,\mathrm{M_\odot}$ in the high-resolution region to observations. A perfect feedback prescription that is able to replicate all the observables in the universe does not exist. If the prescription is tuned to certain observables, it might fail to reproduce others, which then requires further iterations to the feedback implementation (e.g. \citet{2018MNRAS.473.4077P}).

{\parskip 2em} The main difference between setups is the conditions for star formation, and this is reflected in the best values of the feedback parameters we find. In Setup 2, with more relaxed star formation criteria, $f_*$ is high, and $\epsilon$ is low as compared to Setup 1. In Setup 2, star particles form with ease, of lower mass but have a larger quantity. In order to match the same observed value of $f_\mathrm{s}$ with Setup 1, we use a higher value of $f_*$, creating star particles with higher mass. However, since we demand a good agreement with the observed $f_\mathrm{d}$, we have to lower the feedback energy efficiency from these higher mass star particles. This adjustment results in a lower $\epsilon$ as compared to Setup 1. Therefore, combining the values of feedback parameters with the star formation criteria, we show the self-consistent characteristics of the feedback processes. 

{\parskip 2em} In Setup 2, the points coalesce around low gas surface density, with more gas being converted to stars due to the higher value of $f_*$ and the relaxed star formation criteria. As a result, in the recovery of the KS relation in both simulation setups, Setup 2 did not perform as well as Setup 1. This inability to obtain an appropriate slope of the KS relation in both setups hints at a fundamental limitation of the \cite{1992ApJ...399L.113C} model. In terms of matching other observed properties, this feedback prescription requires more tuning or parameters.

{\parskip 2em} Looking at the other haloes in the high-resolution region in Setup 2, all but three of the haloes within $10^{10}\,\mathrm{M_\odot}$ and $10^{12}\,\mathrm{M_\odot}$ with the calibrated star formation and feedback prescription are an excellent fit to $f_\mathrm{s}$ and $f_\mathrm{d}$ observed by \citet{2010ApJ...708L..14M}. In comparison to the results from Setup 1, the feedback prescriptions in Setup 2 perhaps suggest universality for haloes within the mass range described. We verify this claim with the zoom simulations of a dwarf galaxy of $10^{10}\,\mathrm{M_\odot}$ with these feedback prescriptions. Through the haloes in the high-resolution region of the MW zoom simulation and the halo in the dwarf galaxy zoom simulation, we demonstrate the insensitivity of our feedback prescription on the mass resolution. However, we have to conduct the same test with much lower mass haloes as well as with different spatial resolutions. On top of the resolution, the universality and robustness of the feedback prescription should also be extended to galaxies with various star formation and merger history. 

{\parskip 2em} As we demonstrate, non-deterministic variance is a cause for concern; more computational resources need to be invested in order to understand, quantify and minimise these effects. Since we do not reproduce all the observational constraints mentioned, there exists the possibility of including more parameters in the feedback model or developing a different model. These should be the focus of future work to improve the feedback prescription in order for the simulated galaxies to better match observations.

\section*{Acknowledgements}

BKO and JAP were supported by the European Research Council under grant number 670193. BKO would like to thank Jose O{\~n}orbe and the TMOX group at the Royal Observatory, Edinburgh for many insightful discussions, and Daniele Sorini's various suggestions to improve the structure of the paper.




\bibliographystyle{mnras}
\bibliography{enzo-guide-to-mw} 



\appendix


\bsp	
\label{lastpage}
\end{document}